\documentclass[aps,prl,superscriptaddress,floatfix,twocolumn,showpacs]{revtex4-1}

\pdfoutput=1

\usepackage{amsmath,dcolumn,bm,amsthm,amsfonts,amssymb,tabularx,subfigure}
\usepackage[colorlinks=true,urlcolor=blue,citecolor=blue,linkcolor=blue]{hyperref}
\usepackage{mathrsfs,times}
\usepackage{bbold,dsfont}
\usepackage{graphicx,graphics,color}
\usepackage{mathtools,nicefrac}
\allowdisplaybreaks

\newcommand{\beq}{\begin{equation}}
\newcommand{\eeq}{\end{equation}}
\newcommand{\bea}{\begin{eqnarray}}
\newcommand{\eea}{\end{eqnarray}}
\newcommand{\widebar}{\overline}
\newcommand{\wtilde}{\widetilde}
\newcommand{\dell}{\Delta\ell}
\newcommand{\R}{\textrm{R}}
\newcommand{\NS}{\textrm{NS}}
\newcommand{\llangle}{\langle\!\langle }
\newcommand{\rrangle}{\rangle\!\rangle }
\newcommand{\Bigrrangle}{\Big\rangle\!\!\Big\rangle }

\begin{document}

\title{Revisiting the Ramond sector of the $\mathcal{N}\!=\!1$ superconformal minimal models}

\author{Chun Chen}
\email[]{chun6@ualberta.ca}
\affiliation{Department of Physics, University of Alberta, Edmonton, Alberta T6G 2E1, Canada}

\author{Joseph Maciejko}
\email[]{maciejko@ualberta.ca}
\affiliation{Department of Physics, University of Alberta, Edmonton, Alberta T6G 2E1, Canada}
\affiliation{Theoretical Physics Institute, University of Alberta, Edmonton, Alberta T6G 2E1, Canada}
\affiliation{Canadian Institute for Advanced Research, Toronto, Ontario M5G 1Z8, Canada}

\date{\today}

\begin{abstract}

Key to the exact solubility of the unitary minimal models in two-dimensional conformal field theory is the organization of their Hilbert space into Verma modules, whereby all eigenstates of the Hamiltonian are obtained by the repeated action of Virasoro lowering operators onto a finite set of highest-weight states. The usual representation-theoretic approach to removing from all modules zero-norm descendant states generated in such a way is based on the assumption that those states form a nested sequence of Verma submodules built upon singular vectors, i.e., descendant highest-weight states. We show that this fundamental assumption breaks down for the Ramond-sector Verma module with highest weight $c/24$ in the even series of $\mathcal{N}\!=\!1$ superconformal minimal models with central charge $c$. To resolve this impasse, we conjecture, and prove at low orders, the existence of a nested sequence of linear-dependence relations that enables us to compute the character of the irreducible $c/24$ module. Based on this character formula, we argue that imposing modular invariance of the torus partition function requires the introduction of a non-null odd-parity Ramond-sector ground state. This symmetrization of the ground-state manifold allows us to uncover a set of conformally invariant boundary conditions not previously discussed and absent in the odd series of superconformal minimal models, and to derive for the first time a complete set of fusion rules for the even series of those models.

\end{abstract}


\maketitle

{\it Introduction.---}Two-dimensional (2D) conformal field theories (CFTs)~\cite{BPZ} play a central role in physics, with applications ranging from string theory and gauge-gravity duality to critical phenomena and strongly correlated electron systems. The presence of additional symmetries besides conformal symmetry can lead to CFTs with a particularly rich mathematical structure. For instance, a minimal generalization of the conformal Virasoro algebra consistent with supersymmetry is the $\mathcal{N}\!=\!1$ super-Virasoro algebra~\cite{ramond1971,neveu1971,*neveu1971b}, given in Eqs.~(\ref{sVir1})--(\ref{sVir3}) below. The $\mathcal{N}\!=\!1$ superconformal minimal models (SMMs)~\cite{FQSPRL,FQS,qiu1986,GKO} are an infinite, discrete series of superconformal field theories (SCFTs) corresponding to unitary, irreducible representations of this algebra, with values $c\!<\!3/2$ of the central charge given by
\begin{align}\label{CentralCharge}
c=\frac{3}{2}\left(1-\frac{8}{m(m+2)}\right),\hspace{5mm}m=2,3,4,\ldots.
\end{align}
As opposed to early proposals for the experimental realization of SMMs at classical multicritical points (e.g., the tricritical Ising model~\cite{FQS}), which requires much fine tuning, advances in condensed matter physics in the past ten years or so suggest several of the models (\ref{CentralCharge}) may more promisingly be realized as {\it bona fide} quantum critical points or even stable quantum critical phases in an increasingly diverse array of platforms, ranging from anyonic spin chains~\cite{Feiguin,gils2009} to boundaries of topological superconductors~\cite{GroverShengVishwanath,Li} and lattice models of interacting Majorana fermions~\cite{RahmaniPRL,*rahmani2015,OBrienFendley,SannomiyaKatsura}.

A first step towards the realization of SMMs in nature by these means is their unambiguous identification in numerical experiments. The entanglement properties of critical (1+1)D quantum many-body systems, to which powerful entanglement-based numerical methods such as the density-matrix renormalization group (DMRG)~\cite{white1992} give direct access, can probe various universal quantities in the underlying CFT and are particularly promising in this regard. For instance, in Ref.~\cite{GroverShengVishwanath} the central charge $c\!=\!7/10$ of the $m\!=\!3$ SMM in the tricritical Ising universality class was determined from a DMRG calculation of the ground-state entanglement entropy~\cite{holzhey1994}. Besides the single value of the central charge, a fuller characterization of the underlying CFT may in principle be achieved by a study of the low-lying entanglement spectrum, which was argued to match the set of scaling dimensions in the boundary CFT~\cite{Lauchli,ohmori2015,cardy2016} for a particular choice of boundary conditions on the entangling surface. This choice of entanglement boundary conditions in turn singles out one among a set of allowed conformally invariant boundary conditions~\cite{cardy1984}, i.e., a pair of boundary (Cardy) states~\cite{Cardy}. By determining numerically how different entanglement boundary conditions affect degeneracies in the low-lying entanglement spectrum, one may in principle directly probe the entire set of fusion rules of the underlying CFT~\cite{Cardy,Verlinde}.

Perhaps surprisingly, the fundamental question of boundary CFT, i.e., the construction of Cardy states --- which additionally yields, via Ref.~\cite{Cardy}, the bulk fusion rules --- has not been satisfactorily settled so far for the even-$m$ series of SMMs. A problem first posed by Ishibashi in his seminal 1989 paper~\cite{Ishibashi}, the construction of Cardy states for the SMMs was completed in the odd-$m$ case by Nepomechie~\cite{Nepomechie}. Subtleties in the even-$m$ case, due to the presence of the Ramond-sector highest weight (HW) $c/24$, were noticed by Apikyan and Sahakyan~\cite{apikyan1998,*apikyan1999}, but the complete set of Cardy states was not obtained. Here we show that the standard assumption of representation theory --- the nested Verma submodule structure of null states --- fails for the $c/24$ module, requiring an alternate approach to the construction of an irreducible module and the computation of its character. We propose as resolution an infinite hierarchy of linear-dependence relations among null states in an auxiliary module whose character is identical to, but easier to compute than, that of the original module. Based on this newly derived character we argue that modular invariance of the torus partition function~\cite{cardy1986} requires the Ramond-sector ground-state manifold to contain states of both fermion parities. We subsequently construct Ishibashi states~\cite{Ishibashi} and solve the Cardy equations~\cite{Cardy}, finding an extra Cardy state, Eq.~(\ref{Cardy_state_even_6}), beyond those discussed in the literature. Inverting the Cardy equations, we correspondingly find two extra fusion rules, Eqs.~(\ref{fusion_rule_1})--(\ref{fusion_rule_2}) and (\ref{fusion_coeff_1}), absent from previous discussions. To our knowledge, this is the first time the full set of Cardy states and fusion rules have been derived for the even-$m$ series of SMMs.

{\it Failure of standard representation theory for the $c/24$ module.}---The $\mathcal{N}\!=\!1$ super-Virasoro algebra is given by
\begin{align}
\left[L_m,L_n\right]&=\left(m-n\right)L_{m+n}+\frac{c}{12}\left(m^3-m\right)\delta_{m+n,0},\label{sVir1} \\
\left\{G_r,G_s\right\}&=2L_{r+s}+\frac{c}{3}\left(r^2-\frac{1}{4}\right)\delta_{r+s,0}, \\
\left[L_m,G_r\right]&=\left(\frac{m}{2}-r\right)G_{m+r},\label{sVir3}
\end{align}
where $L_m$ ($G_r$) are the bosonic (fermionic) Laurent modes of the energy-momentum tensor (supercurrent); $m\!\in\!\mathbb{Z}$ while $r\!\in\!\mathbb{Z}\!+\!\frac{1}{2}$ in the Neveu--Schwarz (NS) sector and $r\!\in\!\mathbb{Z}$ in the Ramond (R) sector. We focus on the holomorphic part of the algebra; identical results are obtained for the antiholomorphic part. The central result in representation theory is the Kac determinant~\cite{FQS}, an expression for the determinant of the Gram matrix in the degenerate subspace at level $\ell$ in a Verma module $V^+$ ($V^-$) with even (odd) fermion parity and HW $h$,
\begin{align}
\textrm{det}\,V^+_0&=1,\ \ \ \ \ \textrm{det}\,V^-_0=h-\frac{c}{24}, \\
\textrm{det}\,V^{\pm}_{\ell>0}&=\left(h-\frac{c}{24}\right)^{\frac{P_{\textrm{R}}(\ell)}{2}}\!\!\prod_{r,s\geqslant 1}\!\!\left(h-h_{r,s}(c)\right)^{P_{\textrm{R}}\left(\ell-\frac{rs}{2}\right)},
\end{align}
where $h_{r,s}(c)$ is a prescribed function of $c$ and the integers $r,s$ which determines the finite set of allowed HWs in the SMMs. We focus on the R sector, where $r\!-\!s$ is odd and the maximal level degeneracy $P_{\textrm{R}}(\ell)$ is obtained from the generating function $\sum_{\ell=0}^{\infty} q^\ell P_{\textrm{R}}(\ell)=\prod_{n=1}^{\infty}(1+q^{n})/(1-q^n)\eqqcolon 1/\varphi_{\textrm{R}}(q)$.

Unitarity and irreducibility require that negative-norm states should be absent from the Verma module and null (i.e., zero-norm) states should be systematically removed. A large class of linearly-independent null states consists of states obtained from the repeated action of lowering operators $L_{-m}$, $G_{-r}$ ($m>0,r\geqslant 0$) onto a singular vector $|\boldsymbol{\chi}\rangle$, that is, a descendant state satisfying the HW condition: $L_n|\boldsymbol{\chi}\rangle=0\ \mbox{for}\ n>0$ and $G_r|\boldsymbol{\chi}\rangle=0\ \mbox{for}\ r>0$. Such states form a null Verma submodule~\cite{BigYellow}. A singular vector is itself necessarily a null state, but the converse is not generally true, as will be seen shortly. For all Virasoro and most super-Virasoro HWs, the first null state dictated by the Kac determinant (at level $\ell=rs/2$) happens to also be a singular vector; the same Kac determinant can thus be used again to find another singular vector in the resulting null Verma submodule, and such a procedure repeats, leading to a nested structure of null Verma submodules which comprises all null states~\cite{Rocha-Caridi}. Based on this nested structure, Kiritsis derived character formulas for NS and R irreducible Verma modules with HWs \emph{not} equal to $c/24$~\cite{Kiritsis}. The $c/24$ HW only appears in the R sector of the even-$m$ series of the SMMs, and we focus on those for the rest of the paper.

The crucial concurrence between first null state and singular vector is however not generically warranted by the Kac determinant, and breaks down for the module with HW $h_\star\!\coloneqq\!h_{\frac{m}{2},\frac{m}{2}+1}(c)\!=\!c/24$. As for other R modules the HW states come in degenerate pairs $|h_\star^\pm\rangle$ of opposite fermion parity where $|h_\star^-\rangle\!=\!G_0|h_\star^+\rangle$, but contrary to other R modules the odd-parity state $|h_\star^-\rangle$ is annihilated by $G_0$ since $G_0^2\!=\!L_0\!-\!\frac{c}{24}$ (and is thus also null). As a result the linear system corresponding to the singular vector condition is in general overdetermined. For example, when $m\!=\!2$, one has $\langle h^+_\star|L_{1}L_{-1}|h^+_\star\rangle\!=\!0$ but $G_{1}L_{-1}|h^+_\star\rangle\!=\!\frac{3}{2}|h^-_\star\rangle\!\neq\!0$, i.e., the first null state $L_{-1}|h^+_\star\rangle$ built from the even-parity HW state $|h^+_\star\rangle$ is not singular. Analogously for $m\!=\!4$, one can prove that no linear combinations of level-$3$ descendants of $|h^+_\star\rangle$ are singular.

{\it Character of the $c/24$ module.---}To resolve this issue we first propose the study of an auxiliary module $\wtilde{V}_\star$, defined by contrast with the original module $V_\star\!\coloneqq\!V(c,h_\star)$ as
\beq\label{defVaux}
V_\star:\ G_0|h^+_\star\rangle=|\mbox{null}\rangle\ \longrightarrow\ \wtilde{V}_\star:\ G_0|\wtilde{h}^+_\star\rangle=0.
\eeq
The entire class of null states built solely on $|h^-_\star\rangle$ has been effectively subsumed into a single representative zero vector so that $|\wtilde{h}^-_\star\rangle$ does not appear and the Verma module is halved, which greatly simplifies the construction of irreducible HW representations. The Kac determinant for $\wtilde{V}_\star$ remains formally unchanged: $\textrm{det}\,\widetilde{V}^+_{\star,0}\!=\!1,\ \textrm{det}\,\wtilde{V}^-_{\star,0}\!=\!0,\ \textrm{det}\,\wtilde{V}^{\pm}_{\star,\ell>0}\!=\!\prod_{r,s\geqslant 1}\!\left(h\!-\!h_{r,s}(c)\boldsymbol\right)^{P_{\textrm{R}}\left(\ell-\frac{rs}{2}\right)}$. Thus irreducible modules constructed from $V_\star$ and $\wtilde{V}_\star$, albeit fundamentally different, necessarily possess the same character.

We first investigate the structure of the reducible auxiliary modules $\wtilde{V}_\star^\pm$ built entirely upon the HW state $|\wtilde{h}^+_{\star}\rangle$. According to the Friedan--Qiu--Shenker (FQS) prescription~\cite{FQS}, a set of linearly independent vectors spanning the level-$\ell$ degenerate subspace is given by $G_{-m_1}G_{-m_2}\cdots L_{-n_1}L_{-n_2}\cdots|\wtilde{h}^+_{\star}\rangle$, where $0<m_1<m_2<\cdots$, $0<n_1\leqslant n_2\leqslant\cdots$, and $\sum_i m_i+\sum_i n_{i}\!=\!\ell$. Imposing $G_0|\wtilde{h}^+_\star\rangle=0$ in Eq.~(\ref{defVaux}) leads to two major differences in the structure of $V_\star^\pm$ and $\wtilde{V}_\star^\pm$:
\begin{enumerate} 
\item[(i)] In contrast to $V^{\pm}_\star$, singular vectors $|\boldsymbol{\wtilde{\chi}}_\star\rangle$ are restored in $\wtilde{V}^{\pm}_\star$ and first appear at levels dictated by the Kac determinant. However, generically $|\boldsymbol{\wtilde{\chi}}_\star\rangle$ is not annihilated by $G_0$ although $|\wtilde{h}^+_\star\rangle$ is.
\item[(ii)] At a given level and for a fixed fermion parity, the set of null descendant states built upon the two degenerate singular vectors $|\boldsymbol{\wtilde{\chi}}_\star\rangle$ and $G_0|\boldsymbol{\wtilde{\chi}}_\star\rangle$ by the FQS prescription is in general linearly dependent; thus null states in $\wtilde{V}_\star^\pm$ do not form Verma submodules.
\end{enumerate}
As will now be argued, the linear-dependence relations among null states evoked in (ii) are organized into an infinite hierarchy that plays a role analogous to that of the nested embedding of null Verma submodules for $h\!\neq\!c/24$ HWs, and allows us to compute the irreducible character of the original $V_\star^\pm$ modules.

The trivial or zeroth echelon in this hierarchy corresponds simply to the first singular vector $|\boldsymbol{\wtilde{\chi}}_\star\rangle$ of a given fermion parity, which appears at level $\ell_0\!\coloneqq\!\frac{m}{4}(\frac{m}{2}+1)$ according to the Kac determinant and can thus be written as a linear combination of level-$\ell_0$ descendants of $|\wtilde{h}_\star^+\rangle$,
\begin{align}\label{ZerothEchelon}
|\boldsymbol{\wtilde{\chi}}_\star\rangle=\boldsymbol{\hat{\mathcal{L}}}_0\left[f^{(0)}_1,\ldots,f^{(0)}_{\frac{1}{2}\!P_\textrm{R}(\dell_0)}\right]|\wtilde{h}_\star^+\rangle,
\end{align}
where $f^{(0)}_1,\ldots,f^{(0)}_{\frac{1}{2}\!P_\textrm{R}(\dell_0)}$ are the coefficients of this linear combination, and we define the $k$th-echelon ($k\!\geqslant\!0$) generalized lowering operator,
\begin{align}
\boldsymbol{\hat{\mathcal{L}}}_k&\left[f^{(k)}_1,\ldots,f^{(k)}_{\frac{1}{2}\!P_\textrm{R}(\dell_k)},g^{(k)}_1,\ldots,g^{(k)}_{\frac{1}{2}\!P_\textrm{R}(\dell_k)}\right] \nonumber \\
&\coloneqq f^{(k)}_1L_{-1}^{\dell_k}+\cdots+f^{(k)}_{\frac{1}{2}\!P_\textrm{R}(\dell_k)}G_{-1}G_{1-\dell_k}\nonumber\\
&\phantom{\coloneqq}+g^{(k)}_1G_{-1}L_{-1}^{\dell_k-1}G_0+\cdots+g^{(k)}_{\frac{1}{2}\!P_\textrm{R}(\dell_k)}G_{-\dell_k}G_0,\label{GLOP}
\end{align}
with $\Delta\ell_k\!\coloneqq\!\ell_k-\ell_{k-1}$ and $\ell_k\!\coloneqq\!(1+k)^2\ell_0$. Eq.~(\ref{GLOP}) is the most general fermion-parity-preserving operator that raises the level of a state by $\Delta\ell_k$, and $\ell_k$ for $k\!\geqslant\!1$ is the level at which higher-level singular vectors appear according to the Kac determinant. The first $\frac{1}{2}P_\textrm{R}(\dell_k)$ terms in Eq.~(\ref{GLOP}) involve bosonic generators while the remaining terms involve fermionic generators times the zero mode operator $G_0$. From Eq.~(\ref{defVaux}) the latter trivially vanish when acting on $|\wtilde{h}_\star^+\rangle$ and are thus excluded from Eq.~(\ref{ZerothEchelon}).

The first echelon in the hierarchy corresponds to the linear dependence of null states built upon $|\boldsymbol{\wtilde{\chi}}_\star\rangle$ and $G_0|\boldsymbol{\wtilde{\chi}}_\star\rangle$, which first appears at level $\ell_1$ and can be expressed as
\begin{align}\label{FirstEchelon}
\boldsymbol{\hat{\mathcal{L}}}_1|\boldsymbol{\wtilde{\chi}}_\star\rangle=0.
\end{align}
The set of $f^{(1)}_i$ and $g^{(1)}_i$ coefficients implicit in Eq.~(\ref{FirstEchelon}) is uniquely determined by the $f^{(0)}_i$ up to an overall multiplicative constant. The second echelon expresses the fact that linear dependence relations generated from (\ref{FirstEchelon}) become themselves linearly dependent at higher levels. In general, the $k$th echelon consists of linear-dependence relations among the linear-dependence relations of the ($k-1$)th-echelon, and can be summarized compactly as
\begin{align}\label{kthEchelon}
\boldsymbol{\hat{\mathcal{L}}}_k \boldsymbol{\hat{\mathcal{L}}}_{k-1}|\ell_{k-2}\rangle=0,\hspace{5mm}k\geqslant 2,
\end{align}
where the coefficients $f_i^{(k)},g_i^{(k)}$, again omitted for simplicity, are uniquely determined from the $f_i^{(k-1)},g_i^{(k-1)}$ up to an overall multiplicative constant, and $|\ell_{k-2}\rangle$ denotes an arbitrary state at level $\ell_{k-2}$. In Ref.~\cite{SuppMat} we substantiate this hierarchy conjecture with explicit calculations for $m\!=\!2$; a calculation for $m\!=\!4$ with a computer algebra system yields analogous results.

The hierarchy of linear-dependence relations embodied in Eqs.~(\ref{FirstEchelon})--(\ref{kthEchelon}) can now be used to calculate the character of the original irreducible $c/24$ module $\mathcal{M}^{\pm}_{\frac{m}{2},\frac{m}{2}+1}$: the number of linearly independent null states to be removed from $\wtilde{V}_\star^\pm$ at each level is reduced by one whenever a linear-dependence relation occurs. Implementing this procedure recursively, one obtains~\cite{SuppMat}:
\begin{align}
\chi_{\textrm{R}}(\mathcal{M}^{\pm}_{\frac{m}{2},\frac{m}{2}+1})&=\frac{1}{2\varphi_{\textrm{R}}(q)}\sum_{n\in\mathbb{Z}}\Bigl(q^{\frac{1}{2}m(m+2)n^2} \nonumber \\
&\hspace{17mm}-q^{\frac{1}{2}m(m+2)(n+\frac{1}{2})^2}\Bigr)\pm\frac{1}{2}. \label{char_sym}
\end{align}

{\em Modular invariance of the torus partition function.}---Since the full SCFT is nonlocal, we now restrict ourselves to the Gliozzi--Scherk--Olive (GSO)-projected spin model~\cite{FQS} with even fermion parity. On a torus, the required invariance of the partition function under modular transformations of the torus constrains which HW representations appear in the theory and how often~\cite{cardy1986}. A modular invariant contribution to the partition function for even $m$ was found to be~\cite{MatsuoYahikozawa,Cappelli,Kastor}
\begin{align}
Z^+&=\sum_{(r,s)\in\Delta_\textrm{NS}}\left(|\chi_{\textrm{NS}}(\mathcal{M}_{r,s})|^2+|\chi_{\widetilde{\textrm{NS}}}(\mathcal{M}_{r,s})|^2\right) \nonumber \\
&+|\chi_{\textrm{R}}^{c/24}(q)|^2+2\!\sum'_{(r,s)\in\Delta_\textrm{R}}\!|\chi_{\textrm{R}}(\mathcal{M}^{+}_{r,s})|^2, \label{torus_partition}
\end{align}
where $\Delta_\textrm{NS}$ ($\Delta_\textrm{R}$) denotes the set of independent HWs in the NS (R) sector, $\chi_{\widetilde{\textrm{NS}}}$ corresponds to the NS character twisted by the insertion of the fermion parity operator, and the primed sum over $\Delta_{\textrm{R}}$ means that the $c/24$ HW $(r\!=\!\frac{m}{2},s\!=\!\frac{m}{2}\!+\!1)$ is excluded. The asymmetry in Eq.~(\ref{torus_partition}) between the latter and other R HWs comes from the fact that the $c/24$ HW occupies the self-symmetric point of the Kac table~\cite{MatsuoYahikozawa,Cappelli,Kastor}. The function $\chi_{\textrm{R}}^{c/24}(q)\!\coloneqq\!\varphi_{\textrm{R}}(q)^{-1}\!\sum_{n\in\mathbb{Z}}\bigl(q^{\frac{1}{2}m(m+2)n^2}-q^{\frac{1}{2}m(m+2)(n+\frac{1}{2})^2}\bigr)$, whose form is highly constrained by modular invariance, should be the character of the $c/24$ module but disagrees with the result (\ref{char_sym}) of an explicit evaluation in representation theory. To resolve this paradox we introduce an additional pair of R HW ground states $|w^{\pm}_\star\rangle$, obeying
\beq
L_0|w^{\pm}_\star\rangle\!=\!\frac{c}{24}|w^{\pm}_\star\rangle,\ \langle w^-_\star|w^-_\star\rangle\!=\!1,\ |w^+_\star\rangle\!=\!G_0|w^-_\star\rangle\!=\!|\mbox{null}\rangle,
\eeq
and giving rise to a second irreducible $c/24$ module built on the non-null HW state $|w_\star^-\rangle$. The level degeneracies of the modules built on $|h_\star^+\rangle$ and $|w_\star^-\rangle$ only differ at level zero, since the parity of the non-null HW state is opposite for both, and the sum of their characters inferred from Eq.~(\ref{char_sym}) yields precisely $\chi_\R^{c/24}(q)$. We thus interpret Eq.~(\ref{torus_partition}) as an off-diagonal modular invariant involving the product of holomorphic and antiholomorphic characters for two distinct $c/24$ modules, in sharp contrast to the diagonal form for $m$ odd. The introduction of an additional R ground state also resolves the arbitrariness in the original definition $|h^-_\star\rangle\!=\!G_0|h^+_\star\rangle$~\cite{FQS}, which could have equally been chosen as $|h^+_\star\rangle\!=\!G_0|h^-_\star\rangle$.

{\em Boundary SCFT and a new Cardy state.}---We now explore the consequences of this symmetrization of the R ground-state manifold on the boundary SCFT. Superconformally invariant boundary states, the Cardy states, can be expanded on a basis of Ishibashi states $|h_\gamma\rrangle$, $\gamma=\pm 1$, which are constructed for each irreducible module and obey the gluing conditions $(L_n\!-\!\widebar{L}_{-n})|h_\gamma\rrangle\!=\!0$ and $(G_r+i\gamma\widebar{G}_{-r})|h_\gamma\rrangle\!=\!0$~\cite{Ishibashi}. For the $c/24$ HW, we now have two sets of Ishibashi states,
\begin{align}
|h^\textrm{R}_{\star,\gamma}\rrangle&=\sum_{{\sf q}}|h_\star,{\sf q}\rangle\otimes U_\gamma|\overline{h_\star,{\sf q}}\rangle, \label{Ishibashi_h} \\
|w^\textrm{R}_{\star,\gamma}\rrangle&=\sum_{{\sf q'}}|w_\star,{\sf q'}\rangle\otimes U_\gamma|\overline{w_\star,{\sf q'}}\rangle, \label{Ishibashi_w}
\end{align}
where $U_\gamma$ is an antiunitary operator that commutes with the holomorphic fermion parity operator $(-1)^F$ and obeys $U_\gamma L_n U_\gamma^{-1}=L_n$, $U_\gamma G_r U_\gamma^{-1}=-i\gamma G_r(-1)^F$. The sums run over a complete set of states in each module, with ${\sf q,q'}$ a set of quantum numbers sufficient to label each state (holomorphic fermion parity, level, and other quantum numbers). An equally valid basis, which facilitates the construction of the Cardy states, is given by the bonding and antibonding combinations of Eqs.~(\ref{Ishibashi_h}) and (\ref{Ishibashi_w}),
\beq
\big|\textstyle\frac{c}{24}^\R_\pm\big\rangle\!\big\rangle\coloneqq|h^\R_{\star,\pm}\rrangle\pm|w^\R_{\star,\pm}\rrangle.
\eeq

Cardy states~\cite{Cardy} $\|\alpha_\gamma\rrangle, \|\beta_{\gamma'}\rrangle$ are defined by the property that the partition function $Z_{\alpha_\gamma\beta_{\gamma'}}$ of the (GSO-projected) SCFT on a cylinder of length $L$ and circumference $R$ can be evaluated in either the open-string or closed-string pictures. In the open-string picture, one has periodic time evolution along the $R$ direction according to a Hamiltonian $H_{\alpha_\gamma\beta_{\gamma'}}^\text{open}=\frac{\pi}{L}\left(L_0-\frac{c}{24}\right)_{\alpha_\gamma\beta_{\gamma'}}$ with boundary conditions $\alpha_\gamma,\beta_{\gamma'}$ along the $L$ direction,
\begin{align}
Z_{\alpha_\gamma\beta_{\gamma'}}^\text{open}(q)&=\frac{1}{2}\underset{\{\textrm{NS}\}}{\mathrm{Tr}}\!\big[e^{-RH_{\alpha_\gamma\beta_{\gamma'}}^\text{open}}\big]+\frac{1}{2}\underset{\{\widetilde{\textrm{NS}}\}}{\mathrm{Tr}}\!\big[(-1)^F e^{-RH_{\alpha_\gamma\beta_{\gamma'}}^\text{open}}\big] \nonumber \\
&+\frac{1}{2}\underset{\{\textrm{R}\}}{\mathrm{Tr}}\big[e^{-RH_{\alpha_\gamma\beta_{\gamma'}}^\text{open}}\big]+\frac{1}{2}\underset{\{\widetilde{\textrm{R}}\}}{\mathrm{Tr}}\big[(-1)^Fe^{-RH_{\alpha_\gamma\beta_{\gamma'}}^\text{open}}\big],
\end{align}
where traces are over holomorphic states only and all four spin structures on the cylinder are considered separately. In the closed-string picture, one has a transition amplitude between Cardy states with finite time evolution along the $L$ direction,
\begin{align}\label{Zclosed}
Z_{\alpha_\gamma\beta_{\gamma'}}^\text{closed}(\wtilde{q})=\llangle\Theta\alpha_\gamma\|e^{-LH^{\textrm{closed}}}\|\beta_{\gamma'}\rrangle,
\end{align}
with Hamiltonian $H^{\textrm{closed}}=\frac{2\pi}{R}\left(L_0+\widebar{L}_0-\frac{c}{12}\right)$, and $\Theta$ is the antiunitary CPT operator. The parameters $q=e^{-\pi R/L}=e^{2\pi i\tau}$ and $\wtilde{q}=e^{-4\pi L/R}=e^{2\pi i\wtilde{\tau}}$ are related by a modular $S$ transformation $\wtilde{\tau}=-1/\tau$. The open-string partition function is evaluated as
\begin{align}\label{Zopen}
Z&_{\alpha_\gamma\beta_{\gamma'}}^\text{open}(q)=\frac{1}{2}\sum_{i\in\Delta_\textrm{NS}}\left(n_{\alpha_\gamma\beta_{\gamma'}}^i\chi_\textrm{NS}^i(q)
+\wtilde{n}_{\alpha_\gamma\beta_{\gamma'}}^i\chi_{\wtilde{\textrm{NS}}}^i(q)\right) \nonumber \\
&+\frac{1}{2}\left(m_{\alpha_\gamma\beta_{\gamma'}}^{h^+_\star}+m_{\alpha_\gamma\beta_{\gamma'}}^{w^-_\star}\right)\chi_\R^{c/24}(q)+\sum'_{i\in\Delta_\R}m_{\alpha_\gamma\beta_{\gamma'}}^i\chi_\R^i(q) \nonumber \\
&+\frac{1}{2}\left(\wtilde{m}_{\alpha_\gamma\beta_{\gamma'}}^{h^+_\star}-\wtilde{m}_{\alpha_\gamma\beta_{\gamma'}}^{w^-_\star}\right) \nonumber \\
&=\frac{1}{2}\sum_{i\in\Delta_\textrm{NS}}\left(n_{\alpha_\gamma\beta_{\gamma'}}^i\chi_\textrm{NS}^i(q)
+\wtilde{n}_{\alpha_\gamma\beta_{\gamma'}}^i\chi_{\wtilde{\textrm{NS}}}^i(q)\right) \nonumber \\
&+\sum_{i\in\Delta_\R}m_{\alpha_\gamma\beta_{\gamma'}}^i\chi_\R^i(q)+\frac{1}{2}\wtilde{m}_{\alpha_\gamma\beta_{\gamma'}}^{c/24},
\end{align}
where the multiplicities $n^i_{\alpha_\gamma\beta_{\gamma'}},\widetilde{n}^i_{\alpha_\gamma\beta_{\gamma'}},m^i_{\alpha_\gamma\beta_{\gamma'}},\widetilde{m}^i_{\alpha_\gamma\beta_{\gamma'}}\!\in\!\mathbb{Z}$ denote how many times the irreducible HW module $i$ appears in the spectrum of the SCFT with boundary conditions $\alpha_\gamma,\beta_{\gamma'}$ for a given choice of spin structure. In the last equality we define $m_{\alpha_\gamma\beta_{\gamma'}}^{h^+_\star}\!+\!m_{\alpha_\gamma\beta_{\gamma'}}^{w^-_\star}\!\eqqcolon\!2m_{\alpha_\gamma\beta_{\gamma'}}^{c/24}$ and $\wtilde{m}_{\alpha_\gamma\beta_{\gamma'}}^{h^+_\star}\!-\!\wtilde{m}_{\alpha_\gamma\beta_{\gamma'}}^{w^-_\star}\!\eqqcolon\!\wtilde{m}_{\alpha_\gamma\beta_{\gamma'}}^{c/24}$, establishing later that $m_{\alpha_\gamma\beta_{\gamma'}}^{c/24}$ is integer. Note that the two independent $c/24$ modules in general contribute a nontrivial constant from the twisted sum in the R sector, which is at the origin of the new Cardy state to be discussed shortly. Conversely, expanding the Cardy states on the basis of Ishibashi states as
\begin{align}\label{CardyExp}
\|\alpha_\gamma\rrangle=\sum_{j\in\Delta_\textrm{NS}}B_{\alpha_\gamma}^{j^\textrm{NS}_\gamma}|j^\textrm{NS}_\gamma\rrangle+\sum_{j\in\Delta_\textrm{R}}B_{\alpha_\gamma}^{j^\textrm{R}_\gamma}|j^\textrm{R}_\gamma\rrangle,
\end{align}
and similarly for $\|\beta_{\gamma'}\rrangle$, where $B_{\alpha_\gamma}^{j^\textrm{NS,R}_\gamma}\!\coloneqq\!\llangle j_\gamma^\textrm{NS,R}\|\alpha_\gamma\rrangle$ are the expansion coefficients, the closed-string partition function (\ref{Zclosed}) can be calculated and expressed in terms of these coefficients and the characters $\chi_\textrm{NS}^j(\wtilde{q})$, $\chi_{\wtilde{\textrm{NS}}}^j(\wtilde{q})$, $\chi_\textrm{R}^j(\wtilde{q})$~\cite{SuppMat}. The aforementioned constant term in Eq.~(\ref{Zopen}) is matched by a corresponding term in Eq.~(\ref{Zclosed}), which arises from
\begin{align}
\big\langle\!\big\langle\Theta\textstyle\frac{c}{24}_\pm^\R\big|e^{-LH^\textrm{closed}}\big|\textstyle\frac{c}{24}_\mp^\R\big\rangle\!\big\rangle=2.
\end{align}
Equating Eqs.~(\ref{Zopen}) and (\ref{Zclosed}), and using the transformation properties of the characters under a modular $S$ transformation~\cite{MatsuoYahikozawa,Cappelli,Kastor}, we find a set of four Cardy equations~\cite{SuppMat},
\begin{align}
\sum_{i\in\Delta_{\textrm{NS}}}n^{i}_{\alpha_{\gamma}\beta_{\gamma'}}S^{[\textrm{NS},\textrm{NS}]}_{ij}=
2(\delta_{\gamma+}\delta_{\gamma'+}+\delta_{\gamma-}\delta_{\gamma'-})
B_{\alpha_\gamma}^{j_\gamma^\textrm{NS}}B_{\beta_{\gamma'}}^{j_{\gamma'}^\textrm{NS}}&, \label{Cardy_eqn_even_1} \\
\sum_{i\in\Delta_{\textrm{R}}}m^{i}_{\alpha_{\gamma}\beta_{\gamma'}}\frac{1}{\lambda_i}S^{[\textrm{R},\widetilde{\textrm{NS}}]}_{ij}=(\delta_{\gamma+}\delta_{\gamma'-}+\delta_{\gamma-}\delta_{\gamma'+})
B_{\alpha_\gamma}^{j_\gamma^\textrm{NS}}B_{\beta_{\gamma'}}^{j_{\gamma'}^\textrm{NS}}&, \\
\sum_{i\in\Delta_{\textrm{NS}}}\widetilde{n}^{i}_{\alpha_{\gamma}\beta_{\gamma'}}
S^{[\widetilde{\textrm{NS}},\textrm{R}]}_{ij}\lambda_j=
4(\delta_{\gamma+}\delta_{\gamma'+}+\delta_{\gamma-}\delta_{\gamma'-})
B_{\alpha_\gamma}^{j_\gamma^\textrm{R}}B_{\beta_{\gamma'}}^{j_{\gamma'}^\textrm{R}}&, \label{Cardy_eqn_even_3} \\
\widetilde{m}^{c/24}_{\alpha_{\gamma}\beta_{\gamma'}}=
4(\delta_{\gamma+}\delta_{\gamma'-}+\delta_{\gamma-}\delta_{\gamma'+})
B_{\alpha_\gamma}^{\frac{c}{24}_\gamma^\textrm{R}}B_{\beta_{\gamma'}}^{\frac{c}{24}_{\gamma'}^\textrm{R}}&, \label{Cardy_eqn_even_4}
\end{align}
where $S$ is the modular $S$-matrix~\cite{Verlinde}, $j$ runs over HWs in the appropriate sector, and we define $\lambda_{c/24}=1$, $\lambda_{j\neq c/24}=\sqrt{2}$.

Adopting the method in Ref.~\cite{Cardy}, the Cardy states can be obtained as solutions to the above equations. For HWs $k,l\in\Delta_{\textrm{NS}}$, we obtain
\begin{align}
{\|}{k}^{{\textrm{NS}}}_{{+}}{\rrangle}&=\frac{1}{\sqrt{2}}\sum_{j\in\Delta_{\textrm{NS}}}\frac{S^{[\textrm{NS},\textrm{NS}]}_{kj}}{\sqrt{S^{[\textrm{NS},\textrm{NS}]}_{0j}}}|j^{\textrm{NS}}_+\rrangle, \label{Cardy_state_even_2} \\
{\|}{l}^{\widetilde{{\NS}}}_{{+}}{\rrangle}&=\sum_{j\in\Delta_{\R}}\frac{\sqrt{\lambda_j}}{2}\frac{S^{[\widetilde{\textrm{NS}},\textrm{R}]}_{lj}}{\sqrt{S^{[\widetilde{\textrm{NS}},\textrm{R}]}_{0j}}}|j^{\R}_{+}\rrangle, \label{Cardy_state_even_4}
\end{align}
while for HWs $d\!\in\!\Delta_\R$, we obtain
\begin{align}
{\|}{d}^{{\R}}_{{-}}{\rrangle}=\sum_{j\in\Delta_{\NS}}\frac{\sqrt{2}}{\lambda_d}\frac{S^{[\textrm{R},\widetilde{\textrm{NS}}]}_{dj}}{\sqrt{S^{[\textrm{NS},\textrm{NS}]}_{0j}}}|j^{\NS}_-\rrangle. \label{Cardy_state_even_5}
\end{align}
Finally, for even $m$ the existence of the self-symmetric $c/24$ HW yields an additional Cardy state,
\begin{align}
{\Big\|}{\textstyle\frac{c}{24}}^{\widetilde{{\R}}}_{{-}}{\Bigrrangle}=\frac{1}{2\sqrt{S^{[\widetilde{\textrm{NS}},\textrm{R}]}_{0,c/24}}}\big|\textstyle\frac{c}{24}^{\R}_{-}\big\rangle\!\big\rangle. \label{Cardy_state_even_6}
\end{align}

{\em Fusion rules.}---Identifying the multiplicities appearing in the open-string partition function (\ref{Zopen}) with the fusion coefficients of the SCFT~\cite{Cardy}, one obtains the complete set of fusion rules for the even-$m$ series of SMMs,
\begin{align}
\left[{\textstyle\frac{c}{24}}^{\widetilde{{\R}}}\right]\times\left[{\textstyle\frac{c}{24}}^{\widetilde{{\R}}}\right]&=\sum_{i\in\Delta_{\NS}}\widetilde{n}^{i}_{{\frac{c}{24}}^{\widetilde{{\R}}},{\frac{c}{24}}^{\widetilde{{\R}}}}\left[{i}^{\widetilde{{\NS}}}\right], \label{fusion_rule_1} \\
\left[{l}^{\widetilde{{\NS}}}\right]\times\left[{\textstyle\frac{c}{24}}^{\widetilde{{\R}}}\right]&=\widetilde{m}^{\frac{c}{24}}_{{l}^{\widetilde{{\NS}}},{\frac{c}{24}}^{\widetilde{{\R}}}}\left[{\textstyle\frac{c}{24}}^{\widetilde{{\R}}}\right], \label{fusion_rule_2} \\
\left[{l}^{\widetilde{{\NS}}}\right]\times\left[{l'}^{\widetilde{{\NS}}}\right]&=\sum_{i\in\Delta_{\NS}}\widetilde{n}^{i}_{{l}^{\widetilde{{\NS}}},{l'}^{\widetilde{{\NS}}}}\left[{i}^{\widetilde{{\NS}}}\right], \\
\left[{d}^{{\R}}\right]\times\left[{d'}^{{\R}}\right]&=\sum_{i\in\Delta_\NS}n^{i}_{{d}^{{\R}},{d'}^{{\R}}}\left[{i}^{{\NS}}\right], \\
\left[{k}^{{\NS}}\right]\times\left[{d}^{{\R}}\right]&=\sum_{i\in\Delta_\R}m^{i}_{{k}^{{\NS}},{d}^{{\R}}}\left[{i}^{{\R}}\right], \\
\left[{k}^{{\NS}}\right]\times\left[{k'}^{{\NS}}\right]&=\sum_{i\in\Delta_\NS}n^{i}_{{k}^{{\NS}},{k'}^{{\NS}}}\left[{i}^{{\NS}}\right],
\end{align}
as well as the corresponding Verlinde formula,
\begin{align}
\widetilde{n}^{i}_{{\frac{c}{24}}^{\widetilde{{\R}}},{\frac{c}{24}}^{\widetilde{{\R}}}}&=\frac{\left(S^{[\widetilde{\NS},\R]}\right)^{-1}_{\frac{c}{24},i}}{S^{[\widetilde{\NS},\R]}_{0,\frac{c}{24}}},\ \ \widetilde{m}^{\frac{c}{24}}_{{l}^{\widetilde{{\NS}}},{\frac{c}{24}}^{\widetilde{{\R}}}}=\frac{S^{[\widetilde{\NS},\R]}_{l,\frac{c}{24}}}{S^{[\widetilde{\NS},\R]}_{0,\frac{c}{24}}},
\label{fusion_coeff_1} \\
\widetilde{n}^{i}_{{l}^{\widetilde{{\NS}}},{l'}^{\widetilde{{\NS}}}}&=\sum_{j\in\Delta_\R}\frac{S^{[\widetilde{\NS},\R]}_{l j} S^{[\widetilde{\NS},\R]}_{l' j} \left(S^{[\widetilde{\NS},\R]}\right)^{-1}_{ji}}{S^{[\widetilde{\NS},\R]}_{0j}}, \\
n^{i}_{{d}^{{\R}},{d'}^{{\R}}}&=\!\!\!\sum_{j\in\Delta_\NS}\!\!\!\frac{4 S^{[\R,\widetilde{\NS}]}_{dj} S^{[\R,\widetilde{\NS}]}_{d'j} \left(S^{[\NS,\NS]}\right)^{-1}_{ji}}{\lambda_d\lambda_{d'} S^{[\NS,\NS]}_{0j}}, \\
m^{i}_{{k}^{{\NS}},{d}^{{\R}}}&=\!\!\!\sum_{j\in\Delta_\NS}\!\!\!\frac{\lambda_i S^{[\NS,\NS]}_{kj} S^{[\R,\widetilde{\NS}]}_{dj}\left(S^{[\R,\widetilde{\NS}]}\right)^{-1}_{ji}}{\lambda_d S^{[\NS,\NS]}_{0j}}, \\
n^{i}_{{k}^{{\NS}},{k'}^{{\NS}}}&=\!\!\!\sum_{j\in\Delta_\NS}\!\!\!\frac{S^{[\NS,\NS]}_{kj} S^{[\NS,\NS]}_{k'j} \left(S^{[\NS,\NS]}\right)^{-1}_{ji}}{S^{[\NS,\NS]}_{0j}}.
\end{align}
We have checked explicitly that all fusion coefficients are integers.

{\it Conclusion.---}In summary, we have shown that the standard representation-theoretic approach to the determination of irreducible characters in CFT fails for the R-sector $c/24$ HW in the even-$m$ series of $\mathcal{N}\!=\!1$ SMMs, and have conjectured an infinite hierarchy of linear-dependence relations that allows us to compute the character of this module. Modular invariance on the torus can be restored with this character provided that the R ground-state manifold is augmented by a non-null state with odd fermion parity, which in turn yields additional bulk fusion channels and superconformally invariant boundary states.

We gratefully acknowledge discussions with T. Creutzig as well as financial support from NSERC grant \#RGPIN-2014-4608, the CRC Program, CIFAR, and the University of Alberta.

\bibliography{SCFT}

\end{document}


\title{Supplemental Material for \lq\lq Revisiting the Ramond sector of the $\mathcal{N}\!=\!1$ superconformal minimal models''}

\author{Chun Chen}
\email[]{chun6@ualberta.ca}
\affiliation{Department of Physics, University of Alberta, Edmonton, Alberta T6G 2E1, Canada}

\author{Joseph Maciejko}
\email[]{maciejko@ualberta.ca}
\affiliation{Department of Physics, University of Alberta, Edmonton, Alberta T6G 2E1, Canada}
\affiliation{Theoretical Physics Institute, University of Alberta, Edmonton, Alberta T6G 2E1, Canada}
\affiliation{Canadian Institute for Advanced Research, Toronto, Ontario M5G 1Z8, Canada}

\date{\today}

\maketitle

\tableofcontents

\section{Linear dependence of null descendant states in the auxiliary Verma module $\wtilde{V}^\pm_\star$}

\subsection{Linear independence of descendant states in reducible Verma modules with $h_\R\neq \frac{c}{24}$} \label{linear_independ}

We first review the linear independence of descendant states in the \emph{reducible} Verma module $V^\pm_{h_\R}$ associated to a generic Ramond highest weight (HW) $h_\R$ \emph{other than} $\frac{c}{24}$. Descendant states are obtained by repeatedly acting with bosonic and fermionic lowering operators on the ground-state HW vectors $|h^\pm_\R\rangle$ according to the Friedan--Qiu--Shenker (FQS) prescription~\cite{FQS}. The spacings between eigenvalues of $L_0$ (i.e., the eigenenergies) are integer valued, and eigenstates corresponding to distinct eigenvalues are orthogonal, thus one only needs to concentrate on the construction of the degenerate subspace associated to a particular $L_0$-eigenvalue $h_\R+\ell$, where $\ell=0,1,2,\ldots$ is the level. Without loss of generality, we focus on the even-parity module $V^+_{h_\R}$, consisting of states with even holomorphic fermion parity; results for the odd-parity module $V^-_{h_\R}$ can be obtained straightforwardly by a similar procedure. At level $\ell$, the set of linearly independent (but not necessarily orthogonal) descendant states is given by the union of the two sets of vectors described below.

The first set, built upon $|h^+_\R\rangle$, comprises vectors of the form
\beq
G_{-m_1}G_{-m_2}\cdots L_{-n_1}L_{-n_2}\cdots|h^+_{\R}\rangle, \label{states_h+}
\eeq
where $0<m_1<m_2<\ldots,$ $0<n_1\leqslant n_2\leqslant\ldots,$ $\sum_i m_i+\sum_i n_{i}=\ell$, and the number of fermionic generators in each vector should be even. There are in total $\frac{1}{2}P_\R(\ell)+\frac{1}{2}\delta_{\ell,0}$ states in this first set. For later reference, in Table~\ref{table:PR} we give the Ramond-sector level degeneracy $P_\R(\ell)$ for the first few levels $\ell$, which is calculable from its generating function $\sum_{\ell=0}^\infty t^\ell P_\R(\ell)=\prod_{k=1}^\infty(1+t^k)/(1-t^k)$~\cite{FQS}. 

When $\ell>0$, there exists a second set of vectors, built upon $|h^-_\R\rangle=G_0|h^+_\R\rangle$,
\beq
G_{-m'_1}G_{-m'_2}\cdots L_{-n'_1}L_{-n'_2}\cdots|h^-_{\R}\rangle, \label{states_h-}
\eeq
where $0<m'_1<m'_2<\ldots,$ $0<n'_1\leqslant n'_2\leqslant\ldots,$ $\sum_i m'_i+\sum_i n'_{i}=\ell$, and the number of fermionic generators in Eq.~(\ref{states_h-}) should be odd, as the holomorphic fermion parity of $|h^-_{\R}\rangle$ is odd. There are in total $\frac{1}{2}P_\R(\ell)$ states ($\ell>0$) in this second set.

Because $|h^+_\R\rangle$ and $|h^-_\R\rangle$ are linearly independent, and neither of them is zero, the above FQS prescription ensures that all the states given in Eqs.~(\ref{states_h+}) and (\ref{states_h-}) are linearly independent, such that the dimension of the $\ell$-th eigenspace in $V^+_{h_\R}$ is $P_\R(\ell)$. More generally, for any \emph{descendant} HW state at level $n>0$, i.e., a singular vector $|\boldsymbol{\chi}^+_n\rangle$ in $V^+_{h_\R}$, one can analogously generate a (even-parity or odd-parity) null Verma submodule by applying lowering operators to the degenerate states $|\boldsymbol{\chi}^+_n\rangle$ and $G_0|\boldsymbol{\chi}^+_n\rangle$ following the same FQS prescription. However, as will be revealed below in Sec.~\ref{existence_linear_depend}, the auxiliary Verma module $\wtilde{V}^\pm_\star$ with self-symmetric HW $h_\star=\frac{c}{24}$ has the property that null descendant states built upon singular vectors generically become linearly \emph{dependent}.

\begin{table}
\begin{center}
\begin{tabular}{ |c|c|c|c|c|c|c|c|c|c| }
\hline
 & $\ell=0$ & $\ell=1$ & $\ell=2$ & $\ell=3$ & $\ell=4$ & $\ell=5$ & $\ell=6$ & $\ell=7$ & $\ell=8$ \\ 
\hline
$P_\R(\ell)$ & $1$ & $2$ & $4$ & $8$ & $14$ & $24$ & $40$ & $64$ & $100$ \\ 
\hline
\end{tabular}
\caption{Ramond-sector level degeneracy for the first nine levels.}
\label{table:PR}
\end{center}
\end{table}

\begin{table}
\begin{center}
\begin{tabular}{ |c|c|c||c|c|c| }
\hline
\multicolumn{3}{|c||}{Vector content of $\wtilde{V}^+_\star$} & \multicolumn{3}{c|}{Vector content of $\wtilde{V}^-_\star$} \\
\hline
level & dimension & independent vectors & level & dimension & independent vectors \\
\hline
$\ell=0$ & $1$ & $|\wtilde{h}^+_{\star}\rangle$ & $\ell=0$ & $0$ & $\varnothing$ \\ 
\hline
$\ell=1$ & $1$ & $L_{-1}|\wtilde{h}^+_{\star}\rangle$ & $\ell=1$ & $1$ & $G_{-1}|\wtilde{h}^+_{\star}\rangle$ \\ 
\hline
$\ell=2$ & $2$ & $L_{-1}^2|\wtilde{h}^+_{\star}\rangle$ & $\ell=2$ & $2$ & $G_{-1}L_{-1}|\wtilde{h}^+_{\star}\rangle$ \\ 
 &  & $L_{-2}|\wtilde{h}^+_{\star}\rangle$ & &  & $G_{-2}|\wtilde{h}^+_{\star}\rangle$ \\ 
\hline
$\ell=3$ & $4$ & $L_{-1}^3|\wtilde{h}^+_{\star}\rangle$ & $\ell=3$ & $4$ & $G_{-1}L_{-1}^2|\wtilde{h}^+_{\star}\rangle$ \\ 
 &  & $L_{-1}L_{-2}|\wtilde{h}^+_{\star}\rangle$ &  &  & $G_{-1}L_{-2}|\wtilde{h}^+_{\star}\rangle$ \\
 &  & $L_{-3}|\wtilde{h}^+_{\star}\rangle$ &  &  & $G_{-2}L_{-1}|\wtilde{h}^+_{\star}\rangle$ \\
 &  & $G_{-1}G_{-2}|\wtilde{h}^+_{\star}\rangle$ &  &  & $G_{-3}|\wtilde{h}^+_{\star}\rangle$ \\ 
\hline
$\ell=4$ & $\boldsymbol{7}$ & $L_{-1}^4|\wtilde{h}^+_{\star}\rangle$ & $\ell=4$ & $7$ & $G_{-1}L_{-1}^3|\wtilde{h}^+_{\star}\rangle$ \\ 
 &  & $L_{-1}^2L_{-2}|\wtilde{h}^+_{\star}\rangle$ &  &  & $G_{-1}L_{-1}L_{-2}|\wtilde{h}^+_{\star}\rangle$ \\
 &  & $L_{-2}^2|\wtilde{h}^+_{\star}\rangle$ &  &  & $G_{-1}L_{-3}|\wtilde{h}^+_{\star}\rangle$ \\
 &  & $L_{-1}L_{-3}|\wtilde{h}^+_{\star}\rangle$ &  &  & $G_{-2}L_{-1}^2|\wtilde{h}^+_{\star}\rangle$ \\ 
 &  & $L_{-4}|\wtilde{h}^+_{\star}\rangle$ &  &  & $G_{-2}L_{-2}|\wtilde{h}^+_{\star}\rangle$ \\ 
 &  & $G_{-1}G_{-2}L_{-1}|\wtilde{h}^+_{\star}\rangle$ &  &  & $G_{-3}L_{-1}|\wtilde{h}^+_{\star}\rangle$ \\ 
 &  & $G_{-1}G_{-3}|\wtilde{h}^+_{\star}\rangle$ &  &  & $G_{-4}|\wtilde{h}^+_{\star}\rangle$ \\  
\hline
\end{tabular}
\caption{Lowest five eigenspaces for the auxiliary Verma modules $\wtilde{V}^\pm_\star$.}
\label{table:ModVerma}
\end{center}
\end{table}

\subsection{Explicit structure of the auxiliary Verma module $\wtilde{V}^\pm_\star$} \label{existence}

As a solution to the problem of computing the character of the irreducible module $\mathcal{M}_{\frac{m}{2},\frac{m}{2}+1}^\pm$ associated with the self-symmetric Ramond HW $h_\star=h_{\frac{m}{2},\frac{m}{2}+1}=\frac{c}{24}$, we suggest in the main text the study of the auxiliary (reducible) module $\wtilde{V}_\star$,
\beq
V_\star:\ G_0|h^+_{\star}\rangle=|\mbox{null}\rangle\ \ \ \longrightarrow\ \ \ \wtilde{V}_\star:\ G_0|\wtilde{h}^+_{\star}\rangle=0.
\eeq
In order to demonstrate the implementation of the aforementioned FQS construction for this specific situation, in Table~\ref{table:ModVerma} we explicitly write down the linearly-independent vectors that span the lowest five eigenspaces for both the even-parity ($\wtilde{V}^+_\star$) and odd-parity ($\wtilde{V}^-_\star$) modules. Due to the fact that $G_0|\wtilde{h}^+_{\star}\rangle=0$, unlike in the original (reducible) module $V_\star^\pm$ there is no nonzero odd-parity HW ground state $|\wtilde{h}^-_{\star}\rangle$. Thus according to the FQS prescription in Eqs.~(\ref{states_h+}) and (\ref{states_h-}), one readily deduces that the dimension of the level-$\ell$ eigenspace is $\frac{1}{2}P_\R(\ell)\pm\frac{1}{2}\delta_{\ell,0}$ for the respective Verma modules $\wtilde{V}^\pm_\star$. Clearly, all the states generated by applying the FQS prescription to the HW \emph{ground state} (i.e., not those built upon singular vectors) are manifestly linearly independent, such that the resulting state space obeys the definition of a Verma module.

\subsection{Linear dependence of null descendant states in $\wtilde{V}^\pm_\star$} \label{existence_linear_depend}

In Sec.~\ref{linear_independ}, we claimed that for a singular vector $|\boldsymbol{\wtilde{\chi}}^+_n\rangle$ at level $n>0$ in $\wtilde{V}^+_\star$, the space of (null) descendant states built upon the degenerate singular vectors $|\boldsymbol{\wtilde{\chi}}^+_n\rangle$ and $G_0|\boldsymbol{\wtilde{\chi}}^+_n\rangle$ using the FQS prescription as in Eqs.~(\ref{states_h+}) and (\ref{states_h-}) is in general no longer a Verma module, because of the existence of linear-dependence relations among null descendant states at higher levels.

To demonstrate this assertion, we proceed in two steps. First, in this subsection we illustrate via an example the existence of such linear-dependence relations. In Sec.~\ref{sec:SII}, we argue that analogous linear-dependence relations among null descendant states built upon singular vectors appearing at successively higher levels can be organized into a hierarchical structure, superficially similar to the embedding structure of null Verma submodules in standard representation theory~\cite{Rocha-Caridi} but with an entirely different interpretation, which allows for the evaluation of the character of $\mathcal{M}_{\frac{m}{2},\frac{m}{2}+1}^\pm$ through the consistent subtraction of all linearly-independent null states, level by level.

\begin{table}[b]
\begin{center}
\begin{tabular}{ |c|c|c| }
\hline
level & number & null states \\
\hline
$\ell=0$ & $0$ & $\varnothing$ \\ 
\hline
$\ell=1$ & $1$ & $|\boldsymbol{\wtilde{\chi}}^+_1\rangle$ \\ 
\hline
$\ell=2$ & $2$ & $L_{-1}|\boldsymbol{\wtilde{\chi}}^+_1\rangle$ \\ 
 &  & $G_{-1}G_0|\boldsymbol{\wtilde{\chi}}^+_1\rangle$ \\ 
\hline
$\ell=3$ & $4$ & $L_{-1}^2|\boldsymbol{\wtilde{\chi}}^+_1\rangle$ \\ 
 &  & $L_{-2}|\boldsymbol{\wtilde{\chi}}^+_1\rangle$ \\
 &  & $G_{-1}L_{-1}G_0|\boldsymbol{\wtilde{\chi}}^+_1\rangle$ \\
 &  & $G_{-2}G_{0}|\boldsymbol{\wtilde{\chi}}^+_1\rangle$ \\ 
\hline
$\ell=4$ & $\boldsymbol{8}$ & $L_{-1}^3|\boldsymbol{\wtilde{\chi}}^+_1\rangle$ \\ 
 &  & $L_{-1}L_{-2}|\boldsymbol{\wtilde{\chi}}^+_1\rangle$ \\
 &  & $L_{-3}|\boldsymbol{\wtilde{\chi}}^+_1\rangle$ \\
 &  & $G_{-1}G_{-2}|\boldsymbol{\wtilde{\chi}}^+_1\rangle$ \\ 
 &  & $G_{-1}L_{-1}^2G_0|\boldsymbol{\wtilde{\chi}}^+_1\rangle$ \\ 
 &  & $G_{-1}L_{-2}G_0|\boldsymbol{\wtilde{\chi}}^+_1\rangle$ \\  
 &  & $G_{-2}L_{-1}G_0|\boldsymbol{\wtilde{\chi}}^+_1\rangle$ \\ 
 &  & $G_{-3}G_0|\boldsymbol{\wtilde{\chi}}^+_1\rangle$ \\  
\hline
\end{tabular}
\caption{Even-parity null states in the $m=2$ minimal model built upon $|\boldsymbol{\wtilde{\chi}}^+_1\rangle$ and $G_0|\boldsymbol{\wtilde{\chi}}^+_1\rangle$ from the FQS prescription.}
\label{table:null}
\end{center}
\end{table}

We first consider the simplest possible example, the minimal model with $m=2$ and $c=0$, in which all states at level $\ell>0$ are automatically null. While trivial from a physics standpoint, this example will allow us to uncover a structure which we conjecture persists in the nontrivial models with even $m>2$ and $c>0$. According to the Kac determinant for the modules $\wtilde{V}^\pm_\star$,
\beq
\textrm{det}\,\widetilde{V}^+_{\star,0}=1,\ \ \ \textrm{det}\,\wtilde{V}^-_{\star,0}=0,\ \ \ \textrm{det}\,\wtilde{V}^{\pm}_{\star,\ell>0}=\prod_{r,s\geqslant 1}\boldsymbol{\left(}h-h_{r,s}(c)\boldsymbol{\right)}^{P_{\textrm{R}}\left(\ell-\frac{rs}{2}\right)},
\eeq
the first singular vector appears at level $\frac{rs}{2}=1$, because for $m=2$ the corresponding HW is $h_{1,2}=\frac{c}{24}=0$. Indeed, one finds
\beq
L_{1}L_{-1}|\wtilde{h}^+_{\star}\rangle=\frac{c}{12}|\wtilde{h}^+_{\star}\rangle=0\ \ \ \ \ \mbox{and}\ \ \ \ \ G_{1}L_{-1}|\wtilde{h}^+_{\star}\rangle=\frac{3}{2}G_0|\wtilde{h}^+_{\star}\rangle=0,
\eeq
therefore, the states $|\boldsymbol{\wtilde{\chi}}^+_1\rangle\coloneqq L_{-1}|\wtilde{h}^+_{\star}\rangle$ and $G_0|\boldsymbol{\wtilde{\chi}}^+_1\rangle$ are the first singular vectors in $\wtilde{V}^+_\star$ and $\wtilde{V}^-_\star$, respectively~\cite{note}. In view of the fact that $|\boldsymbol{\wtilde{\chi}}^+_1\rangle$ and $G_0|\boldsymbol{\wtilde{\chi}}^+_1\rangle$ are descendant HW states, a common expectation~\cite{BigYellow} would be that applying the FQS prescription to such states will generate a complete, linearly independent set of null states. Following this line of reasoning, in Table~\ref{table:null} we list all the (even-parity) descendant null states generated in such a way, up to level $4$. 

At level $4$, there are $8$ null states built upon $|\boldsymbol{\wtilde{\chi}}^+_1\rangle$ and $G_0|\boldsymbol{\wtilde{\chi}}^+_1\rangle$ as per the FQS construction; however, we recall from Table~\ref{table:ModVerma} that the maximally allowed number of linearly independent null vectors at level $4$ is $7$. (Recall that for $m=2$, all excited states have zero norm. Also, singular vectors are necessarily null states, and applying any lowering operator to a null state yields another null state.) Therefore, we have provided the first concrete example that illustrates the breakdown of the commonly expected linear independence of the FQS prescription when applied to the \emph{descendant} HW states (i.e., singular vectors) in $\wtilde{V}^\pm_\star$. The explicit linear-dependence relation among those $8$ descendant null states is given by:
\beq
\left(-\frac{1}{2}L_{-1}L_{-2}+\frac{7}{2}L_{-3}-2G_{-1}G_{-2}+G_{-1}L_{-1}^{2}G_0-3G_{-2}L_{-1}G_0+\frac{3}{4}G_{-3}G_0\right)|\boldsymbol{\wtilde{\chi}}^+_1\rangle\!=\!0. \label{1st_round_m_2}
\eeq

\section{Hierarchical structure of linear-dependence relations}\label{sec:SII}

Based on an example, we describe the conjectured hierarchical structure of linear-dependence relations obeyed by descendant null states at successively higher levels in the auxiliary Verma modules $\wtilde{V}^\pm_\star$.

\subsection{The zeroth echelon: the first singular vector}

We first establish some notation. Restricting our attention to the even-parity case, for the $\mathcal{N}=1$ superconformal minimal model with generic even integer $m\geqslant2$ the first singular vector $|\boldsymbol{\wtilde{\chi}}_{\frac{m}{2},\frac{m}{2}+1}\rangle$ in the auxiliary Verma module $\wtilde{V}^+_\star$ appears at level 
\beq
\ell_0\coloneqq\frac{m}{4}\left(\frac{m}{2}+1\right).
\eeq
For later use, we also define the quantity
\beq
\ell_k\coloneqq(1+k)^2\ell_0,\hspace{5mm}k\geqslant -1,
\eeq
which corresponds to the level at which higher-level singular vectors appear, according to the Kac determinant.

The first singular vector can thus be written as a linear combination of the $\frac{1}{2}P_\textrm{R}(\ell_0)$ even-parity states generated at level $\ell_0$ by the FQS prescription applied to the (unique) HW state $|\wtilde{h}^+_{\star}\rangle$,
\begin{align}\label{S9}
|\boldsymbol{\wtilde{\chi}}_{\frac{m}{2},\frac{m}{2}+1}\rangle=\boldsymbol{\left(}f^{(0)}_1L_{-1}^{\ell_0}+\cdots+f^{(0)}_{\frac{1}{2}\!P_\textrm{R}(\ell_0)}G_{-1}G_{1-\ell_0}\boldsymbol{\right)}|\wtilde{h}^+_\star\rangle,
\end{align}
where $f_1^{(0)},\ldots,f^{(0)}_{\frac{1}{2}P_\textrm{R}(\ell_0)}$ are rational coefficients, and $\cdots$ denotes all the other products of lowering operators that increase the level by $\ell_0$ and commute with the holomorphic fermion parity operator. For later purposes it will be useful to introduce the {\it $k$-th echelon generalized lowering operator} [Eq.~(9) in the main text],
\begin{align}
\boldsymbol{\hat{\mathcal{L}}}_k&\left[f^{(k)}_1,\ldots,f^{(k)}_{\frac{1}{2}\!P_\textrm{R}(\dell_k)},g^{(k)}_1,\ldots,g^{(k)}_{\frac{1}{2}\!P_\textrm{R}(\dell_k)}\right] \nonumber \\
&\coloneqq f^{(k)}_1L_{-1}^{\dell_k}+\cdots+f^{(k)}_{\frac{1}{2}\!P_\textrm{R}(\dell_k)}G_{-1}G_{1-\dell_k}+g^{(k)}_1G_{-1}L_{-1}^{\dell_k-1}G_0+\cdots+g^{(k)}_{\frac{1}{2}\!P_\textrm{R}(\dell_k)}G_{-\dell_k}G_0, \label{Lgiantoperator}
\end{align}
where we define
\begin{align}
\Delta\ell_k\coloneqq \ell_k-\ell_{k-1},\hspace{5mm}k\geqslant 0,
\end{align}
with $\ell_{-1}=0$. This generalized lowering operator consists of a linear combination, parametrized by the $P_\text{R}(\Delta\ell_k)$ rational coefficients in its argument, of all possible products of lowering operators that increase the level by $\Delta\ell_k$ and commute with the holomorphic fermion parity operator. The first $\frac{1}{2}P_\text{R}(\Delta\ell_k)$ coefficients $f^{(k)}_1,\ldots,f^{(k)}_{\frac{1}{2}\!P_\textrm{R}(\dell_k)}$ multiply products of lowering operators in which the zero mode operator $G_0$ does not appear, whereas the remaining $\frac{1}{2}P_\text{R}(\Delta\ell_k)$ coefficients $g^{(k)}_1,\ldots,g^{(k)}_{\frac{1}{2}\!P_\textrm{R}(\dell_k)}$ multiply products of lowering operators times an additional $G_0$. The products of lowering operators in each term are ordered according to the FQS prescription in Eq.~(\ref{states_h+})--(\ref{states_h-}) [see also Tables~\ref{table:ModVerma}--\ref{table:null}].

Equation~(\ref{S9}) can thus be written in terms of this generalized lowering operator as
\begin{align}
|\boldsymbol{\wtilde{\chi}}_{\frac{m}{2},\frac{m}{2}+1}\rangle\coloneqq& \boldsymbol{\hat{\mathcal{L}}}_0\left[f^{(0)}_1,\ldots,f^{(0)}_{\frac{1}{2}\!P_\textrm{R}(\dell_0)},g^{(0)}_1,\ldots,g^{(0)}_{\frac{1}{2}\!P_\textrm{R}(\dell_0)}\right]|\wtilde{h}^+_\star\rangle.
\end{align}
At this lowest echelon in the hierarchy, the coefficients $g^{(0)}_1,\ldots,g^{(0)}_{\frac{1}{2}\!P_\textrm{R}(\dell_0)}$ are arbitrary since $G_0|\wtilde{h}^+_\star\rangle=0$, and can thus be set to zero, but this will not be the case at higher echelons $k>0$. Since $\Delta\ell_0=\ell_0$, the coefficients $f^{(0)}_1,\ldots,f^{(0)}_{\frac{1}{2}\!P_\textrm{R}(\dell_0)}$ are the same as in Eq.~(\ref{S9}).

We now consider two examples. For the (trivial) $m=2$ minimal model, the first singular vector appears at level $\ell_0=1$, and is given by
\beq
|\boldsymbol{\wtilde{\chi}}_{1,2}\rangle=L_{-1}|\wtilde{h}^+_{1,2}\rangle, \label{singular_m_2}
\eeq
thus in this case, $\dell_0=1$, and the zeroth-echelon generalized lowering operator is given by
\begin{align}
\boldsymbol{\hat{\mathcal{L}}}_0\left[f^{(0)}_1,g^{(0)}_1\right]=f^{(0)}_1L_{-1}
+\left(f^{(0)}_{1}+2g^{(0)}_1\right)G_{-1}G_{0}.
\end{align}  
Since $G_0|\wtilde{h}^+_{1,2}\rangle=0$, one obtains
\beq
\boldsymbol{\hat{\mathcal{L}}}_0\left[f^{(0)}_1,g^{(0)}_1\right]|\wtilde{h}^+_{1,2}\rangle=f^{(0)}_1L_{-1}|\wtilde{h}^+_{1,2}\rangle.
\eeq
Comparing with Eq.~(\ref{singular_m_2}), we find $f^{(0)}_1=1$, and the spurious coefficient $g^{(0)}_1$ is arbitrary, as advertised.

Analogously, for the (nontrivial) $m=4$ minimal model, we find that the first singular vector, appearing at level $\ell_0=3$, can be written as
\beq
|\boldsymbol{\wtilde{\chi}}_{2,3}\rangle=\left((L_{-1})^3-\frac{25}{12}L_{-1}L_{-2}+\frac{31}{9}L_{-3}-\frac{23}{24}G_{-1}G_{-2}\right)|\wtilde{h}^+_{2,3}\rangle. \label{singular_m_4}
\eeq
In this case, one has $\dell_0=3$, and the zeroth-echelon generalized lowering operator is given by
\begin{align}
\boldsymbol{\hat{\mathcal{L}}}_0
\left[f^{(0)}_1,\ldots,f^{(0)}_4,g^{(0)}_1,\ldots,g^{(0)}_4\right]&=
f^{(0)}_1L_{-1}^3+f^{(0)}_{2}L_{-1}L_{-2}+f^{(0)}_{3}L_{-3}+f^{(0)}_{4}G_{-1}G_{-2} \nonumber \\
&\phantom{=}+g^{(0)}_1G_{-1}L_{-1}^{2}G_0+g^{(0)}_2G_{-1}L_{-2}G_0+g^{(0)}_3G_{-2}L_{-1}G_0+g^{(0)}_{4}G_{-3}G_0,
\end{align}
with $8$ coefficients since $P_\textrm{R}(3)=8$ (see Table~\ref{table:PR}). Comparing to Eq.~(\ref{singular_m_4}) then gives explicitly
\beq
\left\{f^{(0)}_1,f^{(0)}_2,f^{(0)}_3,f^{(0)}_4\right\}=\left\{1,\ -\frac{25}{12},\ \frac{31}{9},\ -\frac{23}{24}\right\},
\eeq
and $g^{(0)}_1,\ldots,g^{(0)}_4$ are arbitrary as explained before. Furthermore, since $|\boldsymbol{\wtilde{\chi}}_{2,3}\rangle$ has zero norm, the $f^{(0)}$ coefficients are defined only up to an overall multiplicative constant. By contrast, and as will be described in detail below, the set of $f^{(k)}$ and $g^{(k)}$ coefficients at higher echelons $k>0$ in the hierarchy is determined up to a \emph{single} overall multiplicative constant by the zeroth-echelon coefficients $\left\{f^{(0)}_1,\ldots,f^{(0)}_{\frac{1}{2}\!P_\textrm{R}(\ell_0)}\right\}$ in a recursive fashion.

\subsection{The first echelon: linear dependence of null descendant states}

To motivate our conjecture of the existence of a hierarchical structure in the set of linear-dependence relations for null states in the auxiliary Verma modules $\wtilde{V}^\pm_\star$, we again look to the $m=2$ minimal model for guidance. We start from the lowest-level singular vector given in Eq.~(\ref{singular_m_2}),
\beq
|\boldsymbol{\wtilde{\chi}}_{1,2}\rangle=L_{-1}|\wtilde{h}^+_{1,2}\rangle
=\boldsymbol{\hat{\mathcal{L}}}_0\left[f^{(0)}_1,g^{(0)}_1\right]|\wtilde{h}^+_{1,2}\rangle,
\eeq
which corresponds to the zeroth echelon in the hierarchy. Next, from Eq.~(\ref{1st_round_m_2}) we observe that accompanying the appearance of the second singular vector at level $\ell_1=4$, there arises a unique linear-dependence condition among the $8$ descendant null states listed in the bottom row of Table~\ref{table:null}, which can be recast in the following suggestive form with the aid of the generalized lowering operator (\ref{Lgiantoperator}),
\begin{align}
&\left(-\frac{1}{2}L_{-1}L_{-2}+\frac{7}{2}L_{-3}-2G_{-1}G_{-2}+G_{-1}L_{-1}^{2}G_0-3G_{-2}L_{-1}G_0+\frac{3}{4}G_{-3}G_0\right)L_{-1}|\wtilde{h}^+_{1,2}\rangle \nonumber \\
&\ \ \ \ \ \ \ \ \ \ \ \ \ \ \ \ \ \ \ \ 
=\boldsymbol{\hat{\mathcal{L}}}_1\left[f^{(1)}_1,\ldots,f^{(1)}_4,g^{(1)}_1,\ldots,g^{(1)}_4\right]
\boldsymbol{\hat{\mathcal{L}}}_0\left[f^{(0)}_1,g^{(0)}_1\right]|\wtilde{h}^+_{1,2}\rangle=0. \label{1st_linear_dependent_L_op}
\end{align}
For $m=2$, the first-echelon generalized lowering operator is given explicitly by
\begin{align}\label{S21}
\boldsymbol{\hat{\mathcal{L}}}_1\left[f^{(1)}_1,\ldots,f^{(1)}_4,g^{(1)}_1,\ldots,g^{(1)}_4\right]&=f^{(1)}_1L_{-1}^3+f^{(1)}_{2}L_{-1}L_{-2}+f^{(1)}_{3}L_{-3}+f^{(1)}_{4}G_{-1}G_{-2} \nonumber \\
&\phantom{=}+g^{(1)}_1G_{-1}L_{-1}^{2}G_0+g^{(1)}_2G_{-1}L_{-2}G_0+g^{(1)}_3G_{-2}L_{-1}G_0+g^{(1)}_{4}G_{-3}G_0,
\end{align}
with $8$ coefficients since $\Delta\ell_1=\ell_1-\ell_0=3$ and $P_\textrm{R}(3)=8$. Inserting Eq.~(\ref{S21}) into Eq.~(\ref{1st_linear_dependent_L_op}) and solving for those $8$ coefficients, we obtain the following solution,
\beq
\left\{f^{(1)}_1,f^{(1)}_2,f^{(1)}_3,f^{(1)}_4,g^{(1)}_1,g^{(1)}_2,g^{(1)}_3,g^{(1)}_4\right\}=\left\{0,\ -\frac{1}{2},\ \frac{7}{2},\ -2,\ 1,\ 0,\ -3,\ \frac{3}{4}\right\}. \label{L_1_coffs}
\eeq
It can be verified that with the input of a nonzero $f^{(0)}_1$, this solution is unique up to an overall multiplicative factor, and thus fully determines the first-echelon generalized lowering operator $\boldsymbol{\hat{\mathcal{L}}}_1$ (up to an overall factor). As we argue below, this is the first echelon in an infinite recursive hierarchy: the first singular vector $|\boldsymbol{\wtilde{\chi}}_{\frac{m}{2},\frac{m}{2}+1}\rangle$ determines the form of $\boldsymbol{\hat{\mathcal{L}}}_0$, which determines the form of $\boldsymbol{\hat{\mathcal{L}}}_1$, which determines the form of $\boldsymbol{\hat{\mathcal{L}}}_2$, and so on and so forth at all echelons $k\geqslant 0$.

\subsection{The second echelon: linear dependence of linear-dependence relations}

Now, based on the first linear-dependence relation (\ref{1st_linear_dependent_L_op}) among the null descendant states built upon $|\boldsymbol{\wtilde{\chi}}_{1,2}\rangle$ and $G_0|\boldsymbol{\wtilde{\chi}}_{1,2}\rangle$ at level $4$ in the $m=2$ minimal model, one can generate linear-dependence relations among higher-level null descendant states built upon the same singular vectors by applying the FQS prescription. For example, at level $5$, one would obtain two linear-dependence relations among the $P_\textrm{R}(5-1)=14$ (see Table~\ref{table:PR}) descendant null states built upon $|\boldsymbol{\wtilde{\chi}}_{1,2}\rangle$ and $G_0|\boldsymbol{\wtilde{\chi}}_{1,2}\rangle$, derived by applying $L_{-1}$ and $G_{-1}G_0$ to both sides of Eq.~(\ref{1st_linear_dependent_L_op}):
\begin{align}
L_{-1}\boldsymbol{\hat{\mathcal{L}}}_1\left[f^{(1)}_1,\ldots,f^{(1)}_4,g^1_{(1)},\ldots,g^{(1)}_4\boldsymbol\right]
|\boldsymbol{\wtilde{\chi}}_{1,2}\rangle&=0, \\
G_{-1}G_0\boldsymbol{\hat{\mathcal{L}}}_1\left[f^{(1)}_1,\ldots,f^{(1)}_4,g^{(1)}_1,\ldots,g^{(1)}_4\right]
|\boldsymbol{\wtilde{\chi}}_{1,2}\rangle&=0.
\end{align}
At level $6$, one can similarly generate $4$ linear-dependence relations,
\begin{align}
L_{-1}^2\boldsymbol{\hat{\mathcal{L}}}_1\left[f^{(1)}_1,\ldots,f^{(1)}_4,g^{(1)}_1,\ldots,g^{(1)}_4\right]
|\boldsymbol{\wtilde{\chi}}_{1,2}\rangle&=0, \\
L_{-2}\boldsymbol{\hat{\mathcal{L}}}_1\left[f^{(1)}_1,\ldots,f^{(1)}_4,g^{(1)}_1,\ldots,g^{(1)}_4\right]
|\boldsymbol{\wtilde{\chi}}_{1,2}\rangle&=0, \\
G_{-1}L_{-1}G_0\boldsymbol{\hat{\mathcal{L}}}_1\left[f^{(1)}_1,\ldots,f^{(1)}_4,g^{(1)}_1,\ldots,g^{(1)}_4\right]
|\boldsymbol{\wtilde{\chi}}_{1,2}\rangle&=0, \\
G_{-2}G_0\boldsymbol{\hat{\mathcal{L}}}_1\left[f^{(1)}_1,\ldots,f^{(1)}_4,g^{(1)}_1,\ldots,g^{(1)}_4\right]
|\boldsymbol{\wtilde{\chi}}_{1,2}\rangle&=0.
\end{align}
The proliferation of these linear-dependence relations proceeds up until level $\ell_2=9$, where the third singular vector appears.

At level $9$, naively applying the FQS prescription gives a total of $P_\R(\dell_2=5)=24$ linear-dependence relations,
\begin{align}
L_{-1}^5\boldsymbol{\hat{\mathcal{L}}}_1\left[f^{(1)}_1,\ldots,f^{(1)}_4,g^{(1)}_1,\ldots,g^{(1)}_4\right]
|\boldsymbol{\wtilde{\chi}}_{1,2}\rangle&=0, \label{S29}\\
\vdots& \nonumber \\
G_{-5}G_0\boldsymbol{\hat{\mathcal{L}}}_1\left[f^{(1)}_1,\ldots,f^{(1)}_4,g^{(1)}_1,\ldots,g^{(1)}_4\right]
|\boldsymbol{\wtilde{\chi}}_{1,2}\rangle&=0. \label{S30}
\end{align}
However, by contrast with linear-dependence relations appearing at lower levels, the $24$ linear-dependence relations in Eq.~(\ref{S29})--(\ref{S30}) are {\it no longer entirely independent among themselves}. One can prove that at this level, there arises a unique relation describing the linear dependence of these $24$ linear-dependence relations. This \lq\lq second-echelon'' linear-dependence relation can be expressed as
\beq
\boldsymbol{\hat{\mathcal{L}}}_2\left[f^{(2)}_1,\ldots,f^{(2)}_{12},g^{(2)}_1,\ldots,g^{(2)}_{12}\right]\boldsymbol{\hat{\mathcal{L}}}_1\left[f^{(1)}_1,\ldots,f^{(1)}_4,g^{(1)}_1,\ldots,g^{(1)}_4\right]|\ell_0\rangle=0, \label{2nd_hierarchy}
\eeq
for a set of $24$ coefficients $f^{(2)}_1,\ldots,f^{(2)}_{12},g^{(2)}_1,\ldots,g^{(2)}_{12}$ to be given below. Here, $|\ell_0\rangle$ stands for any state at level $\ell_0$, i.e., an arbitrary eigenstate of $L_0$ with eigenvalue $\frac{c}{24}+\ell_0$. In other words, the $24$ operators acting on the level-$\ell_0$ singular vector $|\boldsymbol{\wtilde{\chi}}_{1,2}\rangle$ in Eq.~(\ref{S29})--(\ref{S30}) are not linearly independent.

Knowing the $8$ coefficients $\left\{f^{(1)}_1,\ldots,f^{(1)}_4,g^{(1)}_1,\ldots,g^{(1)}_4\right\}$ in Eq.~(\ref{L_1_coffs}), Eq.~(\ref{2nd_hierarchy}) can be solved for the $24$ coefficients mentioned above. First, according to the definition (\ref{Lgiantoperator}), the ordered set of operators appearing in $\boldsymbol{\hat{\mathcal{L}}}_2$ is given by
\begin{align}
\Bigl\{&L^5_{-1},\ L^3_{-1}L_{-2},\ L_{-1}L^2_{-2},\ L^2_{-1}L_{-3},\ L_{-2}L_{-3},\ L_{-1}L_{-4}, \nonumber \\
&L_{-5},\ G_{-1}G_{-2}L^2_{-1},\ G_{-1}G_{-2}L_{-2},\ G_{-1}G_{-3}L_{-1},\ G_{-2}G_{-3},\ G_{-1}G_{-4}, \nonumber \\
&G_{-1}L^4_{-1}G_0,\ G_{-1}L^2_{-1}L_{-2}G_0,\ G_{-1}L^2_{-2}G_0,\ G_{-1}L_{-1}L_{-3}G_0,\ G_{-1}L_{-4}G_0,\ G_{-2}L^3_{-1}G_0, \nonumber \\
&G_{-2}L_{-1}L_{-2}G_0,\ G_{-2}G_{-3}G_0,\ G_{-3}L^2_{-1}G_0,\ G_{-3}L_{-2}G_0,\ G_{-4}L_{-1}G_0,\ G_{-5}G_0\Bigr\}. \label{L_2_ops}
\end{align}
Following the same order, we obtain the following solution for the $24$ coefficients,
\begin{align}
\left\{f^{(2)}_{1},\ldots,f^{(2)}_{12},g^{(2)}_{1},\ldots,g^{(2)}_{12}\right\}=&\left\{1,\ -\frac{61}{4},\ 0,\ \frac{325}{4},\ -160,\ -\frac{1}{2}, \right. \nonumber \\
&\left.\ \ \frac{151}{4},\ -\frac{1275}{8},\ -\frac{57}{4},\ \frac{1281}{16},\ -\frac{295}{4},\ \frac{2135}{16},\ \right. \nonumber \\
&\left.\ \ -2,\ 38,\ -\frac{915}{8},\ -230,\ 375,\ \frac{9067}{8},\ \right. \nonumber \\
&\left.\ \ -\frac{13575}{4},\ \frac{37}{2},\ 60,\ -\frac{309}{2},\ \frac{3975}{8},\ \frac{195}{8}\right\}. \label{L_2_coeffs}
\end{align}
The solution is unique, up to an overall multiplicative factor; thus the second-echelon generalized lowering operator $\boldsymbol{\hat{\mathcal{L}}}_2$ is uniquely determined (up to an overall factor) by the coefficients appearing in the first-echelon operator $\boldsymbol{\hat{\mathcal{L}}}_1$.

\subsection{The $k$-th echelon: recursive hierarchy of linear-dependence relations}

Our conjecture, motivated by the $m=2$ example just discussed and further substantiated by computerized symbolic calculations for $m=4$ (first echelon) not explicitly presented here, is that this recursive pattern of linear-dependence relations persists at all higher levels for all $\mathcal{N}=1$ superconformal minimal models with even $m\geqslant 2$ and can be summarized by the following hierarchy.

\begin{enumerate} 
\item[(i)] The zeroth echelon in the hierarchy corresponds simply to the first singular vector $|\boldsymbol{\wtilde{\chi}}_{\frac{m}{2},\frac{m}{2}+1}\rangle$, appearing at level $\ell_0$,
\beq
\boldsymbol{\hat{\mathcal{L}}}_0\left[f^{(0)}_1,\ldots,f^{(0)}_{\frac{1}{2}\!P_\textrm{R}(\dell_0)}\right]|\wtilde{h}^+_{\frac{m}{2},\frac{m}{2}+1}\rangle=|\boldsymbol{\wtilde{\chi}}_{\frac{m}{2},\frac{m}{2}+1}\rangle,
\eeq
which uniquely determines the set of coefficients $\left\{f^{(0)}_1,\ldots,f^{(0)}_{\frac{1}{2}\!P_\textrm{R}(\dell_0)}\right\}$ up to an overall multiplicative constant.
\item[(ii)] The first echelon consists of linear-dependence relations among descendant null states built upon $|\boldsymbol{\wtilde{\chi}}_{\frac{m}{2},\frac{m}{2}+1}\rangle$ and $G_0|\boldsymbol{\wtilde{\chi}}_{\frac{m}{2},\frac{m}{2}+1}\rangle$. Such relations first appear at level $\ell_1=4\ell_0$, where the second singular vector also appears. This can be expressed as
\beq\label{FirstEchelon}
\boldsymbol{\hat{\mathcal{L}}}_1\left[f^{(1)}_1,\ldots,f^{(1)}_{\frac{1}{2}\!P_\textrm{R}(\dell_1)},g^{(1)}_1,\ldots,g^{(1)}_{\frac{1}{2}\!P_\textrm{R}(\dell_1)}\right]
\boldsymbol{\hat{\mathcal{L}}}_0\left[f^{(0)}_1,\ldots,f^{(0)}_{\frac{1}{2}\!P_\textrm{R}(\dell_0)}\right]|\wtilde{h}^+_{\frac{m}{2},\frac{m}{2}+1}\rangle=0,
\eeq
where the set of coefficients $\left\{f^{(1)}_1,\ldots,f^{(1)}_{\frac{1}{2}\!P_\textrm{R}(\dell_1)},g^{(1)}_1,\ldots,g^{(1)}_{\frac{1}{2}\!P_\textrm{R}(\dell_1)}\right\}$ is uniquely determined by $\left\{f^{(0)}_1,\ldots,f^{(0)}_{\frac{1}{2}\!P_\textrm{R}(\dell_0)}\right\}$, up to an overall multiplicative constant. 
\item[(iii)] The second echelon consists of linear-dependence relations among the linear-dependence relations of the first echelon. Such relations first appear at level $\ell_2=9\ell_0$, where the third singular vector also appears. This can be expressed as
\beq
\boldsymbol{\hat{\mathcal{L}}}_2\left[f^{(2)}_1,\ldots,f^{(2)}_{\frac{1}{2}\!P_\textrm{R}(\dell_2)},g^{(2)}_1,\ldots,g^{(2)}_{\frac{1}{2}\!P_\textrm{R}(\dell_2)}\right]
\boldsymbol{\hat{\mathcal{L}}}_1\left[f^{(1)}_1,\ldots,f^{(1)}_{\frac{1}{2}\!P_\textrm{R}(\dell_1)},g^{(1)}_1,\ldots,g^{(1)}_{\frac{1}{2}\!P_\textrm{R}(\dell_1)}\right]|\ell_0\rangle=0,
\eeq
where $|\ell_0\rangle$ is an arbitrary state at level $\ell_0$ and the set of coefficients $\left\{f^{(2)}_1,\ldots,f^{(2)}_{\frac{1}{2}\!P_\textrm{R}(\dell_2)},g^{(2)}_1,\ldots,g^{(2)}_{\frac{1}{2}\!P_\textrm{R}(\dell_2)}\right\}$ is uniquely determined by $\left\{f^{(1)}_1,\ldots,f^{(1)}_{\frac{1}{2}\!P_\textrm{R}(\dell_1)},g^{(1)}_1,\ldots,g^{(1)}_{\frac{1}{2}\!P_\textrm{R}(\dell_1)}\right\}$, up to an overall multiplicative constant.   
\item[(iv)] In general, the $k$-th echelon ($k\geqslant 2$) consists of linear-dependence relations among the linear-dependence relations of the $(k-1)$-th echelon. Such relations first appear at level $\ell_{k}=(1+k)^2\ell_0$, where the $(k+1)$-th singular vector also appears. This can be expressed as
\beq\label{kthEchelon}
\boldsymbol{\hat{\mathcal{L}}}_{k}\left[f^{(k)}_1,\ldots,f^{(k)}_{\frac{1}{2}\!P_\textrm{R}(\dell_{k})},g^{(k)}_1,\ldots,g^{(k)}_{\frac{1}{2}\!P_\textrm{R}(\dell_{k})}\right]
\boldsymbol{\hat{\mathcal{L}}}_{k-1}\left[f^{(k-1)}_1,\ldots,f^{(k-1)}_{\frac{1}{2}\!P_\textrm{R}(\dell_{k-1})},g^{(k-1)}_1,\ldots,g^{(k-1)}_{\frac{1}{2}\!P_\textrm{R}(\dell_{k-1})}\right]
|\ell_{k-2}\rangle=0,
\eeq
where $|\ell_{k-2}\rangle$ denotes any arbitrary state at level $\ell_{k-2}$ and the set of coefficients $\left\{f^{(k)}_1,\ldots,f^{(k)}_{\frac{1}{2}\!P_\textrm{R}(\dell_{k})},g^{(k)}_1,\ldots,g^{(k)}_{\frac{1}{2}\!P_\textrm{R}(\dell_{k})}\right\}$ is uniquely determined by $\left\{f^{(k-1)}_1,\ldots,f^{(k-1)}_{\frac{1}{2}\!P_\textrm{R}(\dell_{k-1})},g^{(k-1)}_1,\ldots,g^{(k-1)}_{\frac{1}{2}\!P_\textrm{R}(\dell_{k-1})}\right\}$, up to an overall multiplicative constant. 
\end{enumerate}

Our conjecture is that Eq.~(\ref{kthEchelon}) holds for an arbitrary state at level $\ell_{k-2}$, and can be thus regarded as an intrinsic property of the SCFT. Brute-force symbolic computations, as we have done for low $m$ and $k$, unfortunately cannot be carried out at arbitrarily high orders due to the rapid increase in eigenspace dimensions for increasing $m$. A formal, all-orders proof of the conjecture is beyond the scope of our paper and is left as an open problem in mathematics.

\section{Character formula for the irreducible $c/24$ module}

The existence of linear-dependence relations among null states at a given level reduces the number of linearly independent such states to be removed in the construction of an irreducible Verma module, but if those linear-dependence relations are not themselves linearly independent, this oversubtraction must be compensated. Combined with the resulting procedure of successive oversubtractions and compensations, the conjectured hierarchy described above can be used to derive the formula presented in Eq.~(12) of the main text for the character of the irreducible module $\mathcal{M}_{\frac{m}{2},\frac{m}{2}+1}^\pm$.

We first consider the reducible auxiliary module $\wtilde{V}_\star^\pm$, whose state space is illustrated in Table~\ref{table:ModVerma} for the first few levels. The character is
\begin{align}\label{RedAuxChi}
\chi_\R(\wtilde{V}_\star^\pm)={\mathrm{Tr}}_{\textrm{R}}\ q^{L_0-c/24}
=\frac{1}{2}\pm\frac{1}{2}+\frac{1}{2}\sum_{\ell=1}^\infty P_\R(\ell)q^{h_\star+\ell-c/24}=\pm\frac{1}{2}+\frac{1}{2\varphi_\R(q)}.
\end{align}
The character of the irreducible module is obtained by removing the contribution of null states from the trace. The first pair of singular vectors $|\boldsymbol{\wtilde{\chi}}_{\frac{m}{2},\frac{m}{2}+1}\rangle$ and $G_0|\boldsymbol{\wtilde{\chi}}_{\frac{m}{2},\frac{m}{2}+1}\rangle$ appears at level $\ell_0=\frac{rs}{2}$ with $r=\frac{m}{2}$ and $s=\frac{m}{2}+1$. Using the symmetry property $h_{r,s}+\frac{rs}{2}=h_{m+r,m+2-s}$ one finds those states are HW states of a putative reducible Verma submodule with HW $h_\star+\ell_0=h_{\frac{3m}{2},\frac{m}{2}+1}$. This submodule being null, one should subtract its character $\chi_\R(V_{\frac{3m}{2},\frac{m}{2}+1}^\pm)$ from Eq.~(\ref{RedAuxChi}). However, due to the first-echelon linear-dependence relation (\ref{FirstEchelon}) at level $\ell_1$, not all null states of this submodule are linearly independent; prior to subtracting it from (\ref{RedAuxChi}) one should subtract from $\chi_\R(V_{\frac{3m}{2},\frac{m}{2}+1}^\pm)$ the character of the putative reducible submodule appearing at level $\Delta\ell_1=\frac{rs}{2}$ in $V_{\frac{3m}{2},\frac{m}{2}+1}^\pm$, where now $r=\frac{3m}{2}$, $s=\frac{m}{2}+1$. Using the symmetry property above one finds $h_{\frac{3m}{2},\frac{m}{2}+1}+\Delta\ell_1=h_{\frac{5m}{2},\frac{m}{2}+1}$. Proceeding similarly up the echelons of the hierarchy, one obtains the following expression for the character of the irreducible module which, as discussed before, is equal to that of the original module $\mathcal{M}_{\frac{m}{2},\frac{m}{2}+1}^\pm$,
\begin{align}
\chi_\R(\mathcal{M}_{\frac{m}{2},\frac{m}{2}+1}^\pm)&=\pm\frac{1}{2}+\frac{1}{2\varphi_\R(q)}
-\Bigl(\chi_\R(V_{\frac{3m}{2},\frac{m}{2}+1}^\pm)-\Bigl(\chi_\R(V_{\frac{5m}{2},\frac{m}{2}+1}^\pm)-\Bigl(\cdots\Bigr)\Bigr)\Bigr)\nonumber\\
&=\pm\frac{1}{2}+\frac{1}{2\varphi_\R(q)}+\sum_{k=1}^\infty(-1)^k\chi_\R(V_{(2k+1)\frac{m}{2},\frac{m}{2}+1}^\pm).
\end{align}
Using
\begin{align}\label{chiR}
\chi_\R(V_{r,s}^\pm)={\mathrm{Tr}}_{\textrm{R}}\ q^{L_0-c/24}=\sum_{\ell=0}^\infty P_\R(\ell)q^{h_{r,s}+\ell-c/24}=\frac{q^{h_{r,s}-c/24}}{\varphi_\R(q)},
\end{align}
and
\begin{align}
h_{r,s}-\frac{c}{24}=\frac{[(m+2)r-ms]^2}{8m(m+2)},
\end{align}
valid in the Ramond sector, one easily arrives at Eq.~(12) in the main text after a few algebraic manipulations. The infinite sums can be expressed in terms of Jacobi theta functions~\cite{BigYellow}.

\section{Closed-string partition function and worldsheet duality}

An explicit expression for the closed-string partition function is
\begin{align}\label{ZclosedSupp}
Z_{\alpha_\gamma\beta_{\gamma'}}^\text{closed}(\wtilde{q})&=\sum_{j\in\Delta_\textrm{NS}}\Bigl[
\Bigl(\delta_{\gamma+}\delta_{\gamma'+}B_{\alpha_+}^{j_+^\textrm{NS}}B_{\beta_+}^{j_+^\textrm{NS}}+
\delta_{\gamma-}\delta_{\gamma'-}B_{\alpha_-}^{j_-^\textrm{NS}}B_{\beta_-}^{j_-^\textrm{NS}}\Bigr)\chi_\textrm{NS}^j(\wtilde{q})
+\Bigl(\delta_{\gamma+}\delta_{\gamma'-}B_{\alpha_+}^{j_+^\textrm{NS}}B_{\beta_-}^{j_-^\textrm{NS}}+
\delta_{\gamma-}\delta_{\gamma'+}B_{\alpha_-}^{j_-^\textrm{NS}}B_{\beta_+}^{j_+^\textrm{NS}}\Bigr)\chi_{\wtilde{\textrm{NS}}}^j(\wtilde{q})\Bigr]\nonumber\\
&\phantom{=}+2\sum_{j\in\Delta_\R}\Bigl(\delta_{\gamma+}\delta_{\gamma'+}B_{\alpha_+}^{j_+^\textrm{R}}B_{\beta_+}^{j_+^\textrm{R}}
+\delta_{\gamma-}\delta_{\gamma'-}B_{\alpha_-}^{j_-^\textrm{R}}B_{\beta_-}^{j_-^\textrm{R}}\Bigr)\chi_\R^j(\wtilde{q})
+2\Bigl(\delta_{\gamma+}\delta_{\gamma'-}B_{\alpha_+}^{{\frac{c}{24}}_+^\textrm{R}}B_{\beta_-}^{{\frac{c}{24}}_-^\textrm{R}}
+\delta_{\gamma-}\delta_{\gamma'+}B_{\alpha_-}^{{\frac{c}{24}}_-^\textrm{R}}B_{\beta_+}^{{\frac{c}{24}}_+^\textrm{R}}\Bigr).
\end{align}
Comparing Eq.~(\ref{ZclosedSupp}) with the open-string partition function [Eq.~(20) in the main text], into which one substitutes the modular transformation properties of the characters,
\begin{align}
\chi_\textrm{NS}^i(q)&=\sum_{j\in\Delta_\textrm{NS}}S_{ij}^{\textrm{[NS,NS]}}\chi_\text{NS}^j(\wtilde{q}),\\
\chi_{\wtilde{\textrm{NS}}}^i(q)&=\sum_{j\in\Delta_\R}S_{ij}^{[\wtilde{\textrm{NS}},\textrm{R}]}\lambda_j
\chi_\text{R}^j(\wtilde{q}),\\
\chi_\textrm{R}^i(q)&=\frac{1}{\lambda_i}\sum_{j\in\Delta_\textrm{NS}}S_{ij}^{[\textrm{R},\wtilde{\textrm{NS}}]}
\chi_{\wtilde{\textrm{NS}}}^j(\wtilde{q}),
\end{align}
one obtains the Cardy equations (23)--(26) in the main text. Here, the real symmetric modular $S$ matrix, which squares to the identity matrix, has the following explicit form:
\begin{align}
S^{[\textrm{NS},\textrm{NS}]}_{\rho\sigma,rs}&=\frac{4}{\sqrt{m(m+2)}}\sin{\left(\frac{\pi\rho r}{m}\right)}\sin{\left(\frac{\pi\sigma s}{m+2}\right)}, \label{S_matrix_1_even} \\
S^{[\widetilde{\textrm{NS}},\textrm{R}]}_{\rho\sigma,rs}&=\begin{dcases}
\frac{4}{\sqrt{m(m+2)}}(-1)^{\frac{\rho-\sigma}{2}}\sin{\left(\frac{\pi\rho r}{m}\right)}\sin{\left(\frac{\pi\sigma s}{m+2}\right)},\ \ \ \text{if}\ \ (r,s)\neq\left(m/2,m/2+1\right); \\
\frac{2\sqrt{2}}{\sqrt{m(m+2)}}(-1)^{\frac{\rho-\sigma}{2}}\sin{\left(\frac{\pi\rho r}{m}\right)}\sin{\left(\frac{\pi\sigma s}{m+2}\right)},\ \ \ \text{if}\ \ (r,s)=\left(m/2,m/2+1\right),
\end{dcases} \label{S_matrix_2_even} \\
S^{[\textrm{R},\widetilde{\textrm{NS}}]}_{\rho\sigma,rs}&=\begin{dcases}
\frac{4}{\sqrt{m(m+2)}}(-1)^{\frac{r-s}{2}}\sin{\left(\frac{\pi\rho r}{m}\right)}\sin{\left(\frac{\pi\sigma s}{m+2}\right)},\ \ \ \text{if}\ \ (\rho,\sigma)\neq\left(m/2,m/2+1\right); \\ 
\frac{2\sqrt{2}}{\sqrt{m(m+2)}}(-1)^{\frac{r-s}{2}}\sin{\left(\frac{\pi\rho r}{m}\right)}\sin{\left(\frac{\pi\sigma s}{m+2}\right)},\ \ \ \text{if}\ \ (\rho,\sigma)=\left(m/2,m/2+1\right).
\end{dcases} \label{S_matrix_3_even}
\end{align}

\section{Fusion coefficients for the $m=4$ superconformal minimal model}

In this section, we provide explicit values for the fusion coefficients of the $m=4$ superconformal minimal model with central charge $c=1$, by applying the Verlinde formula given in Eqs.~(37)--(41) of the main text. The finite set of HWs for this model can be arranged into the Kac table (Table~\ref{table:KacTable}). There are $4$ distinct HWs in the NS sector,
\beq
\Delta_\NS=\left\{h_{1,1}=0^\NS;\ h_{2,2}=\frac{1}{16}^\NS;\ h_{3,1}=1^\NS;\ h_{3,3}=\frac{1}{6}^\NS\right\},
\eeq
and $4$ distinct HWs in the R sector,
\beq
\Delta_\R=\left\{h_{2,1}=\frac{3}{8}^\R;\ h_{2,3}=\frac{1}{24}^\R;\ h_{3,2}=\frac{9}{16}^\R;\ h_{3,4}=\frac{1}{16}^\R\right\}.
\eeq
The self-symmetric R HW $c/24=1/24$ occupies the centre of the Kac table and thus appears only once.

\begin{table}
\begin{center}
\setlength{\tabcolsep}{10pt}
\renewcommand{\arraystretch}{1.5}
\begin{tabular}{ |c||c|c|c| }
 \hline
            & \multicolumn{3}{ c |}{$h_{r,s}$} \\
 \hline\hline
 $s=5$ & $1^{\textrm{NS}}$ & $\frac{3}{8}^{\textrm{R}}$ & $0^{\textrm{NS}}$ \\ [0.5em]
 $s=4$ & $\frac{9}{16}^{\textrm{R}}$ & $\frac{1}{16}^{\textrm{NS}}$ & $\frac{1}{16}^{\textrm{R}}$ \\ [0.5em]
 $s=3$ & $\frac{1}{6}^{\textrm{NS}}$ & $\frac{1}{24}^{\textrm{R}}$ & $\frac{1}{6}^{\textrm{NS}}$ \\ [0.5em]
 $s=2$ & $\frac{1}{16}^{\textrm{R}}$ & $\frac{1}{16}^{\textrm{NS}}$ & $\frac{9}{16}^{\textrm{R}}$ \\ [0.5em]
 $s=1$ & $0^{\textrm{NS}}$ & $\frac{3}{8}^{\textrm{R}}$ & $1^{\textrm{NS}}$ \\ [0.5em]
 \hline \hline
            & $r=1$ & $r=2$ & $r=3$ \\
 \hline
\end{tabular}
\caption{Kac table of the $m=4$ superconformal minimal model.}
\label{table:KacTable}
\end{center}
\end{table}

Utilizing the $S$-matrix elements from Eqs.~(\ref{S_matrix_1_even})--(\ref{S_matrix_3_even}), a straightforward calculation using the Verlinde formula then yields the fusion matrices, where the row indices (corresponding to the subscripts of the fusion coefficients) denote the pair of channels to be fused, and the column indices (corresponding to the superscripts of the fusion coefficients) stand for the possible fusion outcomes. We obtain:
\beq
\widetilde{n}^{i}_{{\frac{1}{24}}^{\widetilde{{\R}}},{\frac{1}{24}}^{\widetilde{{\R}}}}=\bordermatrix{
& 0^{\widetilde{{\NS}}} & \frac{1}{16}^{\widetilde{{\NS}}} & 1^{\widetilde{{\NS}}} & \frac{1}{6}^{\widetilde{{\NS}}} \cr
 {\frac{1}{24}}^{\widetilde{{\R}}},{\frac{1}{24}}^{\widetilde{{\R}}} & \mathbf{1} & \mathbf{0} & \mathbf{1} & \mathbf{1} \cr 
 &  &  &  &  \cr },
\eeq
\beq
\widetilde{m}^{{\frac{1}{24}}}_{l^{\widetilde{{\NS}}},{\frac{1}{24}}^{\widetilde{{\R}}}}=\bordermatrix{
& {{\frac{1}{24}}^{\widetilde{{\R}}}} \cr
 0^{\widetilde{{\NS}}},{\frac{1}{24}}^{\widetilde{{\R}}} & \mathbf{1} \cr 
 \frac{1}{16}^{\widetilde{{\NS}}},{\frac{1}{24}}^{\widetilde{{\R}}} & \mathbf{0} \cr
 1^{\widetilde{{\NS}}},{\frac{1}{24}}^{\widetilde{{\R}}} & \mathbf{1} \cr
 \frac{1}{6}^{\widetilde{{\NS}}},{\frac{1}{24}}^{\widetilde{{\R}}} & \mathbf{1} \cr
 & \cr },
\eeq
\beq
\widetilde{n}^{i}_{l^{\widetilde{{\NS}}},{l'}^{\widetilde{{\NS}}}}=\bordermatrix{
& 0^{\widetilde{{\NS}}} & \frac{1}{16}^{\widetilde{{\NS}}} & 1^{\widetilde{{\NS}}} & \frac{1}{6}^{\widetilde{{\NS}}} \cr
 0^{\widetilde{{\NS}}},0^{\widetilde{{\NS}}} & \mathbf{1} & \mathbf{0} & \mathbf{0} & \mathbf{0} \cr
 0^{\widetilde{{\NS}}},\frac{1}{16}^{\widetilde{{\NS}}} & \mathbf{0} & \mathbf{1} & \mathbf{0} & \mathbf{0} \cr
 0^{\widetilde{{\NS}}},1^{\widetilde{{\NS}}} & \mathbf{0} & \mathbf{0} & \mathbf{1} & \mathbf{0} \cr
 0^{\widetilde{{\NS}}},\frac{1}{6}^{\widetilde{{\NS}}} & \mathbf{0} & \mathbf{0} & \mathbf{0} & \mathbf{1} \cr
 \frac{1}{16}^{\widetilde{{\NS}}},0^{\widetilde{{\NS}}} & \mathbf{0} & \mathbf{1} & \mathbf{0} & \mathbf{0} \cr
 \frac{1}{16}^{\widetilde{{\NS}}},\frac{1}{16}^{\widetilde{{\NS}}} & \mathbf{1} & \mathbf{0} & \mathbf{-1} & \mathbf{0} \cr
 \frac{1}{16}^{\widetilde{{\NS}}},1^{\widetilde{{\NS}}} & \mathbf{0} & \mathbf{-1} & \mathbf{0} & \mathbf{0} \cr
 \frac{1}{16}^{\widetilde{{\NS}}},\frac{1}{6}^{\widetilde{{\NS}}} & \mathbf{0} & \mathbf{0} & \mathbf{0} & \mathbf{0} \cr
 1^{\widetilde{{\NS}}},0^{\widetilde{{\NS}}} & \mathbf{0} & \mathbf{0} & \mathbf{1} & \mathbf{0} \cr
 1^{\widetilde{{\NS}}},\frac{1}{16}^{\widetilde{{\NS}}} & \mathbf{0} & \mathbf{-1} & \mathbf{0} & \mathbf{0} \cr
 1^{\widetilde{{\NS}}},1^{\widetilde{{\NS}}} & \mathbf{1} & \mathbf{0} & \mathbf{0} & \mathbf{0} \cr
 1^{\widetilde{{\NS}}},\frac{1}{6}^{\widetilde{{\NS}}} & \mathbf{0} & \mathbf{0} & \mathbf{0} & \mathbf{1} \cr
 \frac{1}{6}^{\widetilde{{\NS}}},0^{\widetilde{{\NS}}} & \mathbf{0} & \mathbf{0} & \mathbf{0} & \mathbf{1} \cr
 \frac{1}{6}^{\widetilde{{\NS}}},\frac{1}{16}^{\widetilde{{\NS}}} & \mathbf{0} & \mathbf{0} & \mathbf{0} & \mathbf{0} \cr
 \frac{1}{6}^{\widetilde{{\NS}}},1^{\widetilde{{\NS}}} & \mathbf{0} & \mathbf{0} & \mathbf{0} & \mathbf{1} \cr
 \frac{1}{6}^{\widetilde{{\NS}}},\frac{1}{6}^{\widetilde{{\NS}}} & \mathbf{1} & \mathbf{0} & \mathbf{1} & \mathbf{-1} \cr 
 &  &  &  &  \cr },
\eeq
\beq
{n}^{i}_{d^{{{\R}}},{d'}^{{{\R}}}}=\bordermatrix{
& 0^{{{\NS}}} & \frac{1}{16}^{{{\NS}}} & 1^{{{\NS}}} & \frac{1}{6}^{{{\NS}}} \cr
 \frac{3}{8}^{{{\R}}},\frac{3}{8}^{{{\R}}} & \mathbf{2} & \mathbf{0} & \mathbf{2} & \mathbf{0} \cr
 \frac{3}{8}^{{{\R}}},\frac{1}{24}^{{{\R}}} & \mathbf{0} & \mathbf{0} & \mathbf{0} & \mathbf{4} \cr
 \frac{3}{8}^{{{\R}}},\frac{9}{16}^{{{\R}}} & \mathbf{0} & \mathbf{2} & \mathbf{0} & \mathbf{0} \cr
 \frac{3}{8}^{{{\R}}},\frac{1}{16}^{{{\R}}} & \mathbf{0} & \mathbf{2} & \mathbf{0} & \mathbf{0} \cr
 \frac{1}{24}^{{{\R}}},\frac{3}{8}^{{{\R}}} & \mathbf{0} & \mathbf{0} & \mathbf{0} & \mathbf{4} \cr
 \frac{1}{24}^{{{\R}}},\frac{1}{24}^{{{\R}}} & \mathbf{4} & \mathbf{0} & \mathbf{4} & \mathbf{4} \cr
 \frac{1}{24}^{{{\R}}},\frac{9}{16}^{{{\R}}} & \mathbf{0} & \mathbf{4} & \mathbf{0} & \mathbf{0} \cr
 \frac{1}{24}^{{{\R}}},\frac{1}{16}^{{{\R}}} & \mathbf{0} & \mathbf{4} & \mathbf{0} & \mathbf{0} \cr
 \frac{9}{16}^{{{\R}}},\frac{3}{8}^{{{\R}}} & \mathbf{0} & \mathbf{2} & \mathbf{0} & \mathbf{0} \cr
 \frac{9}{16}^{{{\R}}},\frac{1}{24}^{{{\R}}} & \mathbf{0} & \mathbf{4} & \mathbf{0} & \mathbf{0} \cr
 \frac{9}{16}^{{{\R}}},\frac{9}{16}^{{{\R}}} & \mathbf{2} & \mathbf{0} & \mathbf{0} & \mathbf{2} \cr
 \frac{9}{16}^{{{\R}}},\frac{1}{16}^{{{\R}}} & \mathbf{0} & \mathbf{0} & \mathbf{2} & \mathbf{2} \cr
 \frac{1}{16}^{{{\R}}},\frac{3}{8}^{{{\R}}} & \mathbf{0} & \mathbf{2} & \mathbf{0} & \mathbf{0} \cr
 \frac{1}{16}^{{{\R}}},\frac{1}{24}^{{{\R}}} & \mathbf{0} & \mathbf{4} & \mathbf{0} & \mathbf{0} \cr
 \frac{1}{16}^{{{\R}}},\frac{9}{16}^{{{\R}}} & \mathbf{0} & \mathbf{0} & \mathbf{2} & \mathbf{2} \cr
 \frac{1}{16}^{{{\R}}},\frac{1}{16}^{{{\R}}} & \mathbf{2} & \mathbf{0} & \mathbf{0} & \mathbf{2} \cr
 &  &  &  &  \cr },
\eeq
\beq
{m}^{i}_{k^{{{\NS}}},{d}^{{{\R}}}}=\bordermatrix{
& \frac{3}{8}^{{{\R}}} & \frac{1}{24}^{{{\R}}} & \frac{9}{16}^{{{\R}}} & \frac{1}{16}^{{{\R}}} \cr
 0^{{{\NS}}},\frac{3}{8}^{{{\R}}} & \mathbf{1} & \mathbf{0} & \mathbf{0} & \mathbf{0} \cr
 0^{{{\NS}}},\frac{1}{24}^{{{\R}}} & \mathbf{0} & \mathbf{1} & \mathbf{0} & \mathbf{0} \cr
 0^{{{\NS}}},\frac{9}{16}^{{{\R}}} & \mathbf{0} & \mathbf{0} & \mathbf{1} & \mathbf{0} \cr
 0^{{{\NS}}},\frac{1}{16}^{{{\R}}} & \mathbf{0} & \mathbf{0} & \mathbf{0} & \mathbf{1} \cr
 \frac{1}{16}^{{{\NS}}},\frac{3}{8}^{{{\R}}} & \mathbf{0} & \mathbf{0} & \mathbf{1} & \mathbf{1} \cr
 \frac{1}{16}^{{{\NS}}},\frac{1}{24}^{{{\R}}} & \mathbf{0} & \mathbf{0} & \mathbf{2} & \mathbf{2} \cr
 \frac{1}{16}^{{{\NS}}},\frac{9}{16}^{{{\R}}} & \mathbf{1} & \mathbf{1} & \mathbf{0} & \mathbf{0} \cr
 \frac{1}{16}^{{{\NS}}},\frac{1}{16}^{{{\R}}} & \mathbf{1} & \mathbf{1} & \mathbf{0} & \mathbf{0} \cr
 1^{{{\NS}}},\frac{3}{8}^{{{\R}}} & \mathbf{1} & \mathbf{0} & \mathbf{0} & \mathbf{0} \cr
 1^{{{\NS}}},\frac{1}{24}^{{{\R}}} & \mathbf{0} & \mathbf{1} & \mathbf{0} & \mathbf{0} \cr
 1^{{{\NS}}},\frac{9}{16}^{{{\R}}} & \mathbf{0} & \mathbf{0} & \mathbf{0} & \mathbf{1} \cr
 1^{{{\NS}}},\frac{1}{16}^{{{\R}}} & \mathbf{0} & \mathbf{0} & \mathbf{1} & \mathbf{0} \cr
 \frac{1}{6}^{{{\NS}}},\frac{3}{8}^{{{\R}}} & \mathbf{0} & \mathbf{1} & \mathbf{0} & \mathbf{0} \cr
 \frac{1}{6}^{{{\NS}}},\frac{1}{24}^{{{\R}}} & \mathbf{2} & \mathbf{1} & \mathbf{0} & \mathbf{0} \cr
 \frac{1}{6}^{{{\NS}}},\frac{9}{16}^{{{\R}}} & \mathbf{0} & \mathbf{0} & \mathbf{1} & \mathbf{1} \cr
 \frac{1}{6}^{{{\NS}}},\frac{1}{16}^{{{\R}}} & \mathbf{0} & \mathbf{0} & \mathbf{1} & \mathbf{1} \cr
 &  &  &  &  \cr },
\eeq
\beq
{n}^{i}_{k^{{{\NS}}},{k'}^{{{\NS}}}}=\bordermatrix{
& 0^{{{\NS}}} & \frac{1}{16}^{{{\NS}}} & 1^{{{\NS}}} & \frac{1}{6}^{{{\NS}}} \cr
 0^{{{\NS}}},0^{{{\NS}}} & \mathbf{1} & \mathbf{0} & \mathbf{0} & \mathbf{0} \cr
 0^{{{\NS}}},\frac{1}{16}^{{{\NS}}} & \mathbf{0} & \mathbf{1} & \mathbf{0} & \mathbf{0} \cr
 0^{{{\NS}}},1^{{{\NS}}} & \mathbf{0} & \mathbf{0} & \mathbf{1} & \mathbf{0} \cr
 0^{{{\NS}}},\frac{1}{6}^{{{\NS}}} & \mathbf{0} & \mathbf{0} & \mathbf{0} & \mathbf{1} \cr
 \frac{1}{16}^{{{\NS}}},0^{{{\NS}}} & \mathbf{0} & \mathbf{1} & \mathbf{0} & \mathbf{0} \cr
 \frac{1}{16}^{{{\NS}}},\frac{1}{16}^{{{\NS}}} & \mathbf{1} & \mathbf{0} & \mathbf{1} & \mathbf{2} \cr
 \frac{1}{16}^{{{\NS}}},1^{{{\NS}}} & \mathbf{0} & \mathbf{1} & \mathbf{0} & \mathbf{0} \cr
 \frac{1}{16}^{{{\NS}}},\frac{1}{6}^{{{\NS}}} & \mathbf{0} & \mathbf{2} & \mathbf{0} & \mathbf{0} \cr
 1^{{{\NS}}},0^{{{\NS}}} & \mathbf{0} & \mathbf{0} & \mathbf{1} & \mathbf{0} \cr
 1^{{{\NS}}},\frac{1}{16}^{{{\NS}}} & \mathbf{0} & \mathbf{1} & \mathbf{0} & \mathbf{0} \cr
 1^{{{\NS}}},1^{{{\NS}}} & \mathbf{1} & \mathbf{0} & \mathbf{0} & \mathbf{0} \cr
 1^{{{\NS}}},\frac{1}{6}^{{{\NS}}} & \mathbf{0} & \mathbf{0} & \mathbf{0} & \mathbf{1} \cr
 \frac{1}{6}^{{{\NS}}},0^{{{\NS}}} & \mathbf{0} & \mathbf{0} & \mathbf{0} & \mathbf{1} \cr
 \frac{1}{6}^{{{\NS}}},\frac{1}{16}^{{{\NS}}} & \mathbf{0} & \mathbf{2} & \mathbf{0} & \mathbf{0} \cr
 \frac{1}{6}^{{{\NS}}},1^{{{\NS}}} & \mathbf{0} & \mathbf{0} & \mathbf{0} & \mathbf{1} \cr
 \frac{1}{6}^{{{\NS}}},\frac{1}{6}^{{{\NS}}} & \mathbf{1} & \mathbf{0} & \mathbf{1} & \mathbf{1} \cr 
 &  &  &  &  \cr }.
\eeq
Note that certain fusion coefficients are negative, which is a consequence of the presence of fermionic degrees of freedom in the theory. In superstring theories, the non-negativity condition of the fusion-rule coefficients has to be relaxed to the domain of all integers due to the fact that the loop diagrams of the spacetime fermions might give rise to a relative sign in the characters of the worldsheet partition function~\cite{Dbrane}.